\documentclass[12pt]{elsarticle}

\usepackage{graphicx}
\usepackage{amsmath}
\usepackage{amsthm}
\usepackage{amsfonts}
\usepackage[brazil]{babel}
\usepackage[utf8]{inputenc}
\usepackage{graphicx}
\usepackage{amsmath}
\usepackage{amsthm}
\usepackage{amsfonts}
\usepackage[utf8]{inputenc}

\newfont{\Mb}{msbm10}
\newcommand{\C}{\mbox{\Mb\symbol{67}}}
\newcommand{\R}{\mbox{\Mb\symbol{82}}}

\newcounter{bla}

\newtheorem{teor}{Theorem}[section]
\newtheorem{cor}{Corollary}[section]
\newtheorem{obs}{Remark}[section]
\newtheorem{defin}{Definition}[section]

\newtheorem{algor}{Algorithm}[section]

\journal{Computer Physics Communications}

\begin{document}

\title{\uppercase{Solving First Order Differential Equations presenting elementary functions}}

\author[uerj]{L.G.S. Duarte}
\ead{lgsduarte@gmail.com.br}

\author[uerj]{L.A.C.P. da Mota\corref{cor1}}
\ead{lacpdamota@uerj.br or lacpdamota@gmail.com}

\author[uerj,imetro]{A.B.M.M.  Queiroz}
\ead{andream.melo@gmail.com}

\cortext[cor1]{Corresponding author {\footnotesize \newline L.G.S. Duarte and L.A.C.P. da Mota wish to thank Funda\c c\~ao de Amparo \`a Pesquisa do Estado do Rio de Janeiro (FAPERJ) for a Research Grant.}}

\address[uerj]{Universidade do Estado do Rio de Janeiro,
{\it Instituto de F\'{\i}sica, Depto. de F\'{\i}sica Te\'orica},
{\it 20559-900 Rio de Janeiro -- RJ, Brazil}}

\address[imetro]{Inmetro - Instituto Nacional de Metrologia, Qualidade e Tecnologia.}

\begin{abstract}
We have already dealt with the problem of solving First Order Differential Equations (1ODEs) presenting elementary functions before in \cite{Noscpc2002,Noscpc2012}. In this present paper,  we have established solid theoretical basis through a relation between the 1ODE we are dealing with and a rational second order ordinary differential equation, presenting a Liouvillian first Integral. Here, we have expanded the results in \cite{Noscpc2019}, where we have establish a theoretical background to deal with rational second order ordinary differential equations (2ODEs) via the $S$-function method. Using this generalisation and other results hereby introduced, we have produced a method to integrate the 1ODE under scrutiny. Our methods and algorithm are capable to deal efficiently with chaotic systems, determining regions of integrability.
\end{abstract}


\maketitle

\newpage
\bigskip
\hspace{1pc}

\noindent


\medskip
\section*{Introduction}
Apart from their mathematical interest, dealing with first order differential equation (1ODEs) has its more direct application. Any physical (or, for that matter, any scientific question) that can be formulate as a relationship between the rate of change of some quantity of interest and the quantity itself can be formulated as solving an 1ODE (at some stage). There are many well known phenomena which fall into this category: Concentration and dilution problems are quickly remembered examples. Population models are another such example. More specifically, of non-linear 1ODEs and, as a final example, Astrophysics \cite{astro}, etc. There are many others and there can be many more lurking around the corner of the many scientific endeavours being pursuit. In this sense, any novel mathematical approach, technique to deal with such 1ODEs are welcome and could prove vital to solve a fundamental question. So, due to this great importance (theoretical and practical) of solving 1ODEs, many methods have been improved in the last decades. Regarding the search for elementary and Liouvillian solutions\footnote{The methods that use the Lie symmetry approach have also been greatly improved. See, for example \cite{Olv,Ibr,BlAn,Sch,Noscpc1997,Noscpc1998,AbGu,GoLe,AdMa,GaBrSe,MuRo,MuRo2,PuSa,Nuc}.}, the methods that use the Darboux-Prelle-Singer approach stand out. In this `direction' we can highlight: \cite{Dar,PrSi,Sin,Chr,CaLl,ChLlPaZh,ChGiGiLl,Nosjcam2005,ChLlPe,ChLlPaWa2,ChLlPaWa3,Zha,FeGi,BoChClWe,ChCo}.

We have been studying the existence of (and methods to search for) elementary and Liouvillian first integrals of vector fields in $\R^2$ since 2002 \cite{Noscpc2002,Noscpc2012,Nosjcam2005,Nosjpa2002-1,Nosjpa2002-2,Nosamc2007}. In particular, we tackled the problem of looking for general solutions for 1ODEs with functions in \cite{Noscpc2002,Noscpc2012}. However, the mathematical basis was not laid out in such a solid way and, in addition, the method did not deal well (it was computationally expensive) with the problem of determining Darboux polynomials in three or more variables. This time, we managed to combine the technique we used in \cite{Noscpc2019}\footnote{In that work we use the (so-called) $S$-function (introduced in \cite{Nosjpa2001} and used by us and other people \cite{Noscpc2019,Nosamc2007,LaRa,ChSeLa,Nosjmp2009,Nosjpa2010,GoPiSe,TPCSL,MCSL,tito1,tito2}) that improves greatly the efficiency on the searching for Liouvillian first integrals of 2ODEs.} with a reasoning very close to the idea used in \cite{Noscpc2002,Noscpc2012}\footnote{The central idea is to `transform' elementary functions into rational functions.}. Our method basically starts with an assumption (the existence of a general solution of a specific type) to arrive at an algebraic result: the existence (or not) of a Liouvillian first integral (of a certain type) for a polynomial vector field in three variables and, in the case of a positive answer, the determination of that first integral (and, consequently, of the 1ODE's general solution). Briefly, we associate the 1ODE (presenting an elementary function) with a polynomial vector field in three variables by doing a simple transformation, and then analyze the (possible) existence of a Liouvillian first integral. If the first integral is found we only have to perform the inverse transformation to obtain the general solution of the 1ODE in implicit form.

\bigskip

\noindent
This paper is organized as follows:

\medskip

\noindent
In the first section, we present some basic concepts involved in the Darboux-Prelle-Singer (DPS) approach and in the $S$-function method.

\medskip

\noindent
In the second section, we show how we can associate a 1ODE with an elementary function to a polynomial vector field in three variables and develop a technique similar to the one we built in \cite{Noscpc2019} (to search for Liouvillian first integrals of rational 2ODEs using the $S$-function method) in order to search for a Liouvillian first integral of the vector field. We use these two steps to propose a procedure that can succeed even in cases where the integrating factors have Darboux polynomials of very high degree. In the end of the section (after a few examples in order to clarify the steps of the procedure), we propose a semi algorithm to deal with solving 1ODEs presenting an elementary function and we describe it in a more formal way.

\medskip

\noindent

\medskip

\noindent
In the third section, we discuss the performance of the method. We begin by presenting a set of 1ODEs that our method solves without problems but is not easily solved by the `traditional' methods.
Next, we show that, for an 1ODE presenting parameters, the method employed can perform an analysis of the integrability region of the 1ODE's parameters: We present an example of a reducible 2ODE presenting parameters that the package can successfully handle.

\medskip

\noindent
Finally, we present our conclusions.

\section{Basic concepts and results}
\label{bcr}

In this section we will set the mathematical basis for presenting our method. In the first subsection we will present the Darboux-Prelle-Singer approach in a nutshell and, in the second, the basics of the $S$-function method.

\subsection{The Darboux-Prelle-Singer approach}
\label{dpsa}

When we talk succinctly about the Darboux-Prelle-Singer approach to seeking Liouvillian first integrals of autonomous systems of polynomial 1ODEs in the plane, we can summarize the central idea in one paragraph:

{\it Find the Darboux polynomials (DPs) associated with the 1ODE system and use them to determine an integrating factor.}

Without exaggeration, the determination of the DPs is the most important and (unfortunately) the most complex part (by far) of the whole process. Thus, with a couple of definitions and results, we can describe (briefly) the DPS approach when applied to vector fields in the plane. To begin, let's define more formally the concepts we referred to.
Consider the following system of polynomial 1ODEs in the plane:

\begin{equation}
\label{sys1}
\left\{ \begin{array}{l}
\dot{x} = f(x,y)\\ [2mm]
\dot{y} = g(x,y),
\end{array} \right.
\end{equation}
where $f$ and $g$ are coprime polynomials in $(x,y)$ and the dot means derivative with respect to a parameter $t$ ($\dot{u} \equiv du/dt$).

\begin{defin}
A function $I(x,y)$ is called a {\bf first integral} of the system {\em (\ref{sys1})} if $I$ is constant over the solutions of {\em (\ref{sys1})}.
\end{defin}

\begin{defin}
The vector field associated with the system {\em (\ref{sys1})},
\begin{equation}
\label{vf1}
X \equiv f(x,y)\,\partial_x+g(x,y)\,\partial_y,
\end{equation}
is called {\bf Darboux operator} associated with {\em (\ref{sys1})}.
\end{defin}

\begin{obs}
If $I(x,y)$ is a {\bf first integral} of the system {\em (\ref{sys1})}, then $X(I) = 0$.
\end{obs}

\begin{defin}
The polynomial $p(x,y) \in \C[x,y]$ is called a {\bf Darboux polynomial} of the vector field $X$ if $X(p) = q\,p$, where $q$ is a polynomial denominated {\bf cofactor}.
\end{defin}

\begin{teor}[Prelle-Singer]
\label{ps}
If the system {\em (\ref{sys1})} presents an elementary first integral $I$, then there exists an integrating factor $R$ for the system {\em (\ref{sys1})} of the form $R = \prod_i {p_i}^{n_i}$, where the $p_i$ are irreducible Darboux polynomials of system {\em (\ref{sys1})} and $n_i$ are rational numbers.
\end{teor}

\noindent
{\it Proof.} For a proof see \cite{PrSi}.

\bigskip

Since $\frac{X(R)}{R} = - \,{\rm {\bf div}}(X) = - (f_x + g_y)$, substituting $R = \prod_i {p_i}^{n_i}$ we obtain (see \cite{Noscpc2002})
\begin{equation}
\label{dpispi}
\sum_i \,n_i\,\frac{X(p_i)}{p_i} = \sum_i \,n_i\,q_i = - (f_x + g_y),
\end{equation}
where the polynomials $q_i$ ($\equiv \frac{X(p_i)}{p_i}$) are called {\em cofactors} of the Darboux polynomials $p_i$. Therefore, a possible method to find an elementary first integral is:

\bigskip

\noindent
{\bf Prelle-Singer method:}(sketch)
\begin{itemize}
\item Determine the DPs $p_i$ associated with the system.
\item Find numbers $n_i$ that satisfy $\sum_i \,n_i\,q_i = - (f_x + g_y)$.
\item Construct the integrating factor $R = \prod_i {p_i}^{n_i}$ and find a first integral $I(x,y)$ by quadratures.
\end{itemize}

\begin{obs}
If the system {\em (\ref{sys1})} presents a non elementary Liouvillian first integral the Prelle-Singer (PS) method needs some modifications (see the Christopher-Singer (CS) method in {\em \cite{Nosjcam2005,Noscpc2007,Nosjde2021}}). However, the main premisse remains the same: determine the DPs of the system.
\end{obs}

\begin{obs}
For vector fields in $\R^3$, things get a little more complicated: first, the integrating factor $R$ is no longer a Jacobi multiplier (as it happens in $\R^2$). Thus, $X (R) / R$ is no longer a known polynomial and depends on a rational function not determined a priori. Second, we need to determine Darboux polynomials in three variables (a very difficult task since even determining them in two variables could be trick).
\end{obs}

\medskip

\subsection{The $S$-function method}
\label{sfm}

The concept of S-function emerged when we generalized the DPS approach to deal with rational second order ordinary differential equations (rational 2ODEs). The idea was to `add' a null 1-form to the 1-form that represented the 2ODE (see \cite{Nosjpa2001,Nosjmp2009}).

In \cite{Noscpc2019}, to avoid the very (computationally costly) calculating DPs of high degree (3 or bigger) we developed a mixed procedure we called `the $S$-function method'. Briefly, the method did not use the computation of Darboux polynomials (in exchange for the resolution of two 1ODEs). The good news is that the $S$-function method proved to be more effective precisely in problems in which the Darboux polynomials had a relatively high degree, being a great alternative to be used in these cases. In this section, we will see what the method consists of and its basic operation.

\subsubsection{Some basic definitions and results}
\label{sbdr}

Consider the rational 2ODE given by
\begin{equation}
\label{2oder1}
z'=\frac{dz}{dx}=\phi(x,y,z)=\frac{M(x,y,z)}{N(x,y,z)},  \,\,(z \equiv y'),
\end{equation}
where $M$ and $N$ are coprime polynomials in $\C[x,y,z]$.

\begin{defin}
A function $I(x,y,z)$ is called {\bf first integral} of the 2ODE {\em (\ref{2oder1})} if $I$ is constant over the solutions of {\em (\ref{2oder1})}.
\end{defin}

\begin{obs} If $I(x,y,z)$ is a first integral of the 2ODE {\em (\ref{2oder1})} then, over the solution curves of {\em (\ref{2oder1})}, the exact 1-form $\omega:=dI=I_x\,dx+I_y\,dy+I_{z}\,dz\,$ is null.
\end{obs}
\noindent
Over the solution curves of (\ref{2oder1}), we have two independent null 1-forms:
\begin{eqnarray}
\label{alfa2}
\alpha & := & \phi\,dx-dz, \\
\label{beta2}
\beta & := & z\,dx-dy.
\end{eqnarray}
So, the 1-form $\omega$ is in the vector space sppaned by the 1-forms $\alpha$ and $\beta$, i.e., we can write $\omega = r_1 (x,y,z)\,\alpha+r_2 (x,y,z)\,\beta$:
\begin{eqnarray}
\label{eqform2edo1}
dI &=& I_x\,dx+I_y\,dy+I_{z}\,dy = r_1\,(\phi\,dx - dz)\,+r_2\,(z\,dx - dy) \nonumber \\
    &=& (r_1\,\phi\,+r_2\,z)dx + (-r_2)dy + (-r_1)dz.
\end{eqnarray}
\noindent
If we call $\,r_1 \equiv R\,$ and $\,\frac{r_2}{r_1} \equiv S\,$, we can write
\begin{equation}
\label{dIrands}
dI = R\,\left[(\phi\,+z\,S)dx - S\,dy - dz\right].
\end{equation}

\noindent
Therefore, we have that $I_x = R\,(\phi\,+z\,S), \, I_y = -R\,S, \,I_z = -R$.

\begin{defin}
\label{fatint1form}
Let $\gamma$ be a 1-form. We say that $R$ is an {\bf integrating factor} for the 1-form $\gamma$ if $R\,\gamma$ is an exact 1-form.
\end{defin}

\begin{defin}
Let $I$ be a first integral of the {\em 2ODE (\ref{2oder1})}. The function defined by $S := I_y/I_z$ is called a \mbox{\boldmath $S$}{\bf -function} associated with the {\em 2ODE} through the first integral $I$.
\end{defin}

\begin{obs}
From the above definitions and in view of {\em (\ref{dIrands})} we can see that $R$ is an integrating factor for the 1-form $\,(\phi\,+z\,S)dx - S\,dy - dz\,$ and $S$ is a $S$-function associated with the {\em 2ODE (\ref{2oder1})} through $I$ .
\end{obs}

\begin{teor}
\label{concom}
Let $\,I(x,y,z)\,$ be a first integral of the {\em 2ODE (\ref{2oder1})}. If $S$ and $R$ are as in {\em (\ref{dIrands})}, then we can write
\begin{eqnarray}
\label{eqproof1r}
&& D_x(R) +R\,(S+\phi_z)=0\,, \\
\label{eqproof2r}
&& S\,D_x(R) + R\,( D_x(S)+\phi_y)=0\,, \\
\label{eqproof3r}
&& -(R_z\,S+R\,S_z) + R_y=0\,,
\end{eqnarray}
where $\,D_x := \partial_x + z\,\partial_y + \phi(x,y,z)\,\partial_z\,$.
\end{teor}

\noindent
{\it Proof:} See \cite{Nosjmp2009}.

\begin{cor}
\label{corS}
Let $S$ be a $S$-function associated with the {\em 2ODE} $\,z'=\phi(x,y,z)$. Then $S$ obeys the following equation:
\begin{equation}
\label{eqS}
D_x(S)=S^2+\phi_{z}\,S-\phi_y\,.
\end{equation}
\end{cor}

\noindent
{\it Proof:} See \cite{Noscpc2019}

\subsubsection{The Associated 1ODEs}
\label{conectSandS}
From (\ref{eqS}) we can see that a $S$-function associated with the rational 2ODE (\ref{2oder1}) satisfies a quasilinear 1PDE in the variables $(x,y,z)$.
Over the solutions of the 2ODE (\ref{2oder1}) we have that $y=y(x)$ and $z=z(x)$ and, therefore, the operator $D_x$ is, formally, ${\frac{d}{dx}}$. So, formally, over the solutions of the 2ODE (\ref{2oder1}) we can write the 1PDE (\ref{eqS}) as a Riccati 1ODE: $ds/dx=s^2+\phi_z\,s-\phi_y$. It is of common knowledge that the transformation $y(x)=-\frac{r'(x)}{f(x)\,r(x)}$ changes the Riccati equation $\,y'(x)=f(x)\,y(x)^2+g(x)\,y(x)+h(x)\,$ into the linear 2ODE $\,r''=(({f'(x)+g(x)\,f(x)})\,r')/{f(x)}-f(x)\,h(x)\,r$. So, with the transformation $s(x)=-w'(x)/w(x)$, the Riccati 1ODE turns (over the solutions of the 2ODE (\ref{2oder1})) into the homogeneous linear 2ODE $\,w''={\phi_z}\,w' + \phi_y\,w$. Therefore, we can use the formal equivalence $D_x\,\sim\,{\frac{d}{dx}}$ to produce a connection between the $S$-functions and the symmetries (written in a particular form) of the 2ODE. Applying the transformation
\begin{equation}
\label{stonu}
S=-\frac{D_x(\nu)}{\nu}
\end{equation}
into equation (\ref{eqS}). We obtain:
\begin{equation}
\label{2PDEnu}
D_x^2(\nu)=\phi_z\,D_x(\nu)+\phi_y\,\nu.
\end{equation}
The equation (\ref{2PDEnu}) is the symmetry condition for $\,\nu\,$ to be an infinitesimal that defines a symmetry generator in the evolutionary form. So, we have:
\begin{teor}
\label{connectSSy}
Let $\nu$ be a function of $(x,y,z)$ such that $[0,\nu]$ defines a symmetry of the {\em 2ODE (\ref{2oder1})} in the evolutionary form, i.e., $X_e := \nu\,\partial_y\,$ generates a symmetry transformation for  {\em (\ref{2oder1})}. Then the function defined by $S=-D_x(\nu)/\nu$ is a $S$-function associated with the {\em 2ODE (\ref{2oder1})}.
\end{teor}
\noindent
{\it Proof:} See \cite{Noscpc2019}

\begin{cor}
\label{connectSyS}
Let $S$ be a $S$-function associated with the {\em 2ODE (\ref{2oder1})}. Then, the function $\nu$ given by
\begin{equation}
\label{nuS}
\nu \equiv {\rm e}^{\int_x (-S)},
\end{equation}
(where $\, \int_x \,\,\mbox{\rm is the inverse operator of}\,\, D_x\,, \mbox{\rm i.e.}, \, \int_x\,D_x = D_x\,\int_x = \mbox{\boldmath $1$}$) defines a symmetry of the {\em 2ODE (\ref{2oder1})} in the evolutionary form.
\end{cor}
\noindent
{\it Proof:} See \cite{Noscpc2019}

\bigskip

The theorem \ref{connectSSy} and corollary \ref{connectSyS} establish a connection between $S$-functions and Lie symmetries. It is this connection that allowed us to avoid the use of Darboux polynomials in the searching for first integrals for the 2ODE (\ref{2oder1}). The main concept is the {\em associated 1ODE}\footnote{This concept was developed in \cite{Nosjmp2009}, page 222.}, which is an 1ODE that has its general solution defined by one of the first integrals of the 2ODE.

\begin{defin}
\label{1odeass}
Let $I$ be a first integral of the {\em 2ODE (\ref{2oder1})} and let $S(x,y,z)$ be the $S$-function associated with {\em (\ref{2oder1})} through $I$. The first order ordinary differential equation defined by
\begin{equation}
\label{1odeassdefs}
\frac{dz}{dy}= - S(x,y,z),
\end{equation}
where $x$ is taken as a parameter, is called {\bf 1ODE$_{\mathbf{[1]}}$ associated} with {\em (\ref{2oder1})} through $I$.
\end{defin}

\begin{teor}
\label{sol1odeass}
Let $I$ be a first integral of the {\em 2ODE (\ref{2oder1})} and let $S(x,y,z)$ be the $S$-function associated with {\em (\ref{2oder1})} through $I$. Then $I(x,y,z)=C$ is a general solution of the {\em 1ODE$_{[1]}$ associated} with {\em (\ref{2oder1})} through $I$.
\end{teor}
\noindent
{\it Proof:} See \cite{Noscpc2019}

\begin{obs}
\label{solnotinv}
The variable $x$ (the independent variable of the {\em 2ODE (\ref{2oder1})}) is just a parameter in the {\em 1ODE (\ref{1odeassdefs})}, so, since any function of $x$ is an invariant for the operator  $D_a \equiv \partial_y - S\,\partial_z$, i.e., $D_a(x) = 0$, the relation between a general solution $H(x,y,z)=K$ of the {\em 1ODE (\ref{1odeassdefs})} and the first integral $I(x,y,z)$ of the {\em 2ODE (\ref{2oder1})} is $\,I(x,y,z)={\cal F}\left(x,H\right)$, such that $D_x(I) = \partial_x({\cal F}) + (H_x + z\,H_y + \phi\,H_z)\,\partial_H ({\cal F})= 0$.
\end{obs}

In a short way, the $S$-function method consists in determining the $S$-function, solving the associated 1ODE and solving the 1ODE that represents the characteristic system of the partial differential equation for ${\cal F}$ (see the remark \ref{solnotinv} just above).

\begin{obs}
Some comments:
\label{s1s2s3}

\begin{itemize}
\item Since the $S$-function is defined as the division of $I_y$ by $I_z$, we can think if the divisions $I_x/I_z$ and $I_x/I_y$ would have any meaning or utility. The answer is yes: We can define the $S$-functions $\{S_1,S_2,S_3\}$ with similar meanings (see \cite{Noscpc2019}).
\item The main advantage of the $S$-function method is that, in general, it is more efficient to determine the $S$-function than computing Darboux polynomials precisely in the cases where they (the DPs) have high degrees.
\item A great advantage of having three $S$-functions is that the difficulty associated with determining each one of them can be very different, i.e., it can be much easier to calculate one of them than the others.
\item for more details (and examples) of the application of the method we recommend that the reader see the reference \cite{Noscpc2019}.
\end{itemize}
\end{obs}

\medskip

\section{A new method to solve 1ODEs with an elementary function}
\label{nms1odewf}

In this section we will present a method to find the general solution of 1ODEs with an elementary function\footnote{This version of the method will be limited to non algebraic elementary functions.}. In the first subsection we will show an equivalence between a certain type of 1ODEs and polynomial vector fields in three variables. In the second, we will show how to use the technique developed in \cite{Noscpc2019} to find a Liouvillian first integral of a polynomial vector field in three variables.

\subsection{1ODEs with functions $\times$ polynomial vector fields}
\label{1odewfpvf}


The method we build is applicable to 1ODEs that can be written in the form
\begin{equation}
\label{1odelemb}
y' = \frac{dy}{dx} = \Phi(x,y,\rho)
\end{equation}
where  $\Phi$ is a rational function of  $(x,y,\rho)$, $\rho$ is a rational function of $(x,y,\theta)$ and $\theta$ is an elementary generator over $\C(x,y)$\footnote{This is a formal way of saying that $\rho \in \C(x,y,\theta)$ is an elementary function of $(x,y)$ (see \cite{Dav}) which can be expressed in terms of a single elementary generator $\theta$, e.g., $\rho = \cos(x)=({\rm e}^{ix} + {\rm e}^{-ix})/2 =(\theta + 1/\theta)/2, \, \theta={\rm e}^{ix}$.}. So, we can write (\ref{1odelemb}) as
\begin{equation}
\label{1odelem}
y' = \frac{dy}{dx} = \phi(x,y,\theta) = \frac{M(x,y,\theta)}{N(x,y,\theta)}
\end{equation}
where $\phi \in K = \C(x,y,\theta)$\footnote{$K = \C(x,y,\theta)$ is the differential field of rational functions in the variables $(x,y,\theta)$.} and $M$ and $N$ are coprime polynomials in $(x,y,\theta)$, i.e., $M,\,N \in \C[x,y,\theta]$\footnote{$\C[x,y,\theta]$ is the differential ring of polynomial functions in the variables $(x,y,\theta)$.}. Furthermore, if $I(x,y,\theta) = c$ represents a general solution of 1ODE (\ref{1odelem}) and $I$ is a Liouvillian function of $(x,y,\theta)$ then for the method to work, as we shall see in what follows, it is sufficient that the partial derivatives of $I$ in relation to $x$, $y$ and $\theta$ can be expressed as

\begin{eqnarray}
I_x &=& R\,P_1 \label{ix}  \\
I_y &=& R\,P_2 \label{iy}  \\
I_{\theta} &=& R\,P_3 \label{ithe}
\end{eqnarray}
where $R$ is an elementary function of $(x,y,\theta)$, $P_1,\,P_2,\,P_3 \in \C[x,y,\theta]$ (i.e., $P_1,\,P_2$ and $P_3$ are polynomials in $(x,y,\theta)$ ) and the derivatives of $\theta$, $\frac{\partial \theta}{\partial x},\,\frac{\partial \theta}{\partial y} \in K=\C(x,y,\theta)$.

\medskip

The main idea is to perform the transformation $z=\theta(x,y)$ that allows the association of 1ODE (\ref{1odelem}) with a polynomial vector field that has $I(x,y,z)$ as a Liouvillian first integral. Let's see how this happens:

\begin{itemize}
\item In first place, since $D_x(I)=0$ where $D_x \equiv \partial_x + \phi\,\partial_y$, we have that
\begin{equation}
D_x(I) = \partial_x(I) + \phi\,\partial_y(I) =\frac{\partial I}{\partial x} + \frac{\partial I}{\partial \theta}\,\frac{\partial \theta}{\partial x} + \phi\,\left(\frac{\partial I}{\partial y} + \frac{\partial I}{\partial \theta}\frac{\partial \theta}{\partial y}\right).
\end{equation}

\item Since the derivatives $\frac{\partial \theta}{\partial x},\,\frac{\partial \theta}{\partial y}$ are in $K=\C(x,y,\theta)$, we can write $D_x(I)$ as
\begin{eqnarray}
D_x(I) &=& \frac{\partial I}{\partial x} +  \phi\,\frac{\partial I}{\partial y} + \left(\frac{\partial \theta}{\partial x} + \phi\,\frac{\partial \theta}{\partial y}\right)\,\frac{\partial I}{\partial \theta}=
\nonumber \\ [2mm]
&=& I_x +  \phi\, I_y + \left(\theta_x + \phi\,\theta_y\right)\,I_\theta, \label{eqitem3}
\end{eqnarray}
where $\theta_x \equiv \frac{\partial \theta}{\partial x}$ and $\theta_y \equiv \frac{\partial \theta}{\partial y}$ are rational functions of $(x,y,\theta)$.

\item From the fact that $\phi$, $\theta_x$ and $\theta_y$ are rational functions of $(x,y,\theta)$, we can multiply $D_x(I)$ by the ${\rm lcm}$ of the denominators of $\phi$, $\theta_x$ and $\theta_y\,\phi$ (which we will denote by $l_d$), obtaining
\begin{equation}
\label{fghnovo}
l_d\,D_x(I) = (f\,\partial_x + g\,\partial_y + h\,\partial_{\theta})\left(I(x,y,\theta)\right)=0,
\end{equation}
where $f,\,g$ and $h$ are polynomials in $(x,y,\theta)$ and
\begin{eqnarray}
\label{deefe}
f &=& l_d,
\\ [2mm]
\label{dege}
g &=& l_d\,\phi,
\\ [2mm]
\label{deaga}
h &=& l_d\,\left(\theta_x + \phi\,\theta_y\right).
\end{eqnarray}

\item If we make the substitution $\theta \rightarrow z$, we have that
\begin{equation}
\label{dii0}
(f(x,y,z)\,\partial_x + g(x,y,z)\,\partial_y + h(x,y,z)\,\partial_z)\left(I(x,y,z)\right)=0.
\end{equation}

\end{itemize}

\medskip

\noindent
Therefore, following these steps we can associate the polynomial vector field
\begin{equation}
\label{vfchi}
\chi \equiv f(x,y,z)\,\partial_x + g(x,y,z)\,\partial_y + h(x,y,z)\,\partial_z
\end{equation}
with the 1ODE (\ref{1odelem}). Moreover, we can associate the first integral $I(x,y,z)$ of $\chi$ (see equation (\ref{dii0})) with the general solution of the 1ODE (\ref{1odelem}) (i.e., $I(x,y,\theta)=c$). So, if we can find a Liouvillian first integral $I(x,y,z)$ for the vector field $\chi$, we will have found the general solution of 1ODE (\ref{1odelem}): we only have to apply the substitution $z \rightarrow \theta$ on $I(x,y,z)=c$.

\subsection{$S$-function method adapted to polynomial vector fields in three variables}
\label{sfmpvf3v}

In this section we will show how we can adapt the method described in \cite{Noscpc2019} ($S$-function method) to polynomial vector fields in three variables. Consider that $I(x,y,z)$ is a Liouvillian first integral of the vector field $\chi \equiv f\,\partial_x+g\,\partial_y+h\,\partial_z$, such that the derivatives $I_x$, $I_y$ and $I_z$ are given, respectively, by $R\,P_1$, $R\,P_2$ and $R\,P_3$ (see the section \ref{1odewfpvf}). Now, supose that $I(x,y,z)$ is a Liouvillian first integral of a (hypothetical) rational 2ODE such that $x$ is the independent variable, $y$ is the dependent variable and $z$ is $dy/dx$. This hypothetical 2ODE would be given by
\begin{equation}
\label{hyp2ode}
z' = \Phi(x,y,z) = - \frac{I_x+z\,I_y}{I_z}= - \frac{P_1+z\,P_2}{P_3} = \frac{M_0(x,y,z)}{N_0(x,y,z)},
\end{equation}
where $M_0$ and $N_0$ are coprime polynomials in $(x,y,z)$.

As we pointted out in section \ref{sfm}, more precisely in remark \ref{s1s2s3}, the $S$-functions $S_1, \, S_2 \, {\rm and} \, S_3$ associated with this 2ODE through the first integral $I$ are given by the following rational functions
\begin{equation}
\label{hypsf}
S_1 = \frac{I_y}{I_z}= \frac{P_2}{P_3}, \,\,\,\,S_2 = \frac{I_x}{I_z}= \frac{P_1}{P_3}, \,\,\,\,S_3 = \frac{I_x}{I_y}= \frac{P_1}{P_2},
\end{equation}
where $S_1, \, S_2 \, {\rm and} \, S_3$ obey the equations
\begin{eqnarray}
\label{s1eq}
 D_x(S_1) &=& {S_1}^2+ \Phi_z\,S_1 - \Phi_y,
 \\ [2mm]
\label{s2eq}
 D_x(S_2) &=& -\frac{1}{z}{S_2}^2+\left(\Phi_z-\frac{\Phi}{z}\right)\,S_2-\Phi_x,
 \\ [2mm]
\label{s3eq}
 D_x(S_3) &=& -\frac{\Phi_y}{\Phi}{S_3}^2+\frac{\Phi_x-z\,\Phi\-y}{\Phi}\,S_3+z\,\Phi_x,
\end{eqnarray}
where $D_x \equiv \partial_x+z\,\partial_y+\Phi\,\partial_z$.
\noindent
Since we do not have the function $\Phi$, we cannot use the equations (\ref{s1eq}), (\ref{s2eq}) and (\ref{s3eq}) to find the $S$-functions $S_1, \, S_2 \, {\rm and} \, S_3$ associated with the hypothetical 2ODE (\ref{hyp2ode}). So, in order to use the technique developed in \cite{Noscpc2019}, we will need some relationship between the (hypothetical) 2ODE and the vector field (\ref{vfchi}). We will establish this relationship using the fact that the 2ODE (\ref{hyp2ode}) and the vector field $\chi$ have in common the first integral $I$. This fact allows us to establish a result that will give us a way to apply the method used in \cite{Noscpc2019} to our problem. Let's begin with the following definition:

\begin{defin}
\label{2ode0}
Let $f\,\partial_x+g\,\partial_y+h\,\partial_z$ be a polynomial vector field in three variables presenting a first integral $I(x,y,z)$ and let $z' = \Phi(x,y,z)$ be a rational 2ODE such that it also has $I(x,y,z)$ as a first integral. We say that the 2ODE and the vector field are {\bf associated through the first integral}~$I$.
\end{defin}

\begin{obs}
\label{six2odes}
Note that, since the status of the variables $(x,y,z)$ in the vector field $f\,\partial_x+g\,\partial_y+h\,\partial_z$ are the same, there are, in principle, six distinct 2ODEs associated with it:
\begin{eqnarray}
\label{Phi1}
\Phi_1 &=& - \frac{I_x+z\,I_y}{I_z}, \\ [2mm]
\label{Phi2}
\Phi_2 &=& - \frac{I_y+z\,I_x}{I_z}, \\ [2mm]
\label{Phi3}
\Phi_3 &=& - \frac{I_x+y\,I_z}{I_y}, \\ [2mm]
\label{Phi4}
\Phi_4 &=& - \frac{I_z+y\,I_x}{I_y}, \\ [2mm]
\label{Phi5}
\Phi_5 &=& - \frac{I_y+x\,I_z}{I_x}, \\ [2mm]
\label{Phi6}
\Phi_6 &=& - \frac{I_z+x\,I_y}{I_x}.
\end{eqnarray}
This will be useful in the future process of finding the first integral $I$ through the $S$-function method (see sections \ref{api} and \ref{tpalg}).
\end{obs}

\begin{teor}
\label{s1vfode}
Let $\chi$, defined by {\em (\ref{vfchi})}, be a polynomial vector field presenting a Liouvillian first integral $I$ and let the rational 2ODE defined by {\em (\ref{hyp2ode})} be associated with the vector field $\chi$ through $I$.  If their derivatives are given by
\begin{eqnarray}
I_x &=& R\,P_1 \label{ix1},  \\
I_y &=& R\,P_2 \label{iy1},  \\
I_z &=& R\,P_3 \label{iz1},
\end{eqnarray}
where $R$ is a Darboux function of $(x,y,z)$ and $P_1,\,P_2,\,P_3 \in \C[x,y,z]$ then the $S$-function $S_1$ associated with the rational 2ODE {\em (\ref{hyp2ode})} is given by
\begin{equation}
\label{s12ode1}
S_1 = \frac{M_0\,f - N_0\,h}{N_0\,(g-z\,f)},
\end{equation}
where $g-z\,f \neq 0$.
\end{teor}

\noindent
{\bf Proof of Theorem \ref{s1vfode}:} From the hypotheses we have that

\begin{eqnarray}
\label{dosi}
D_0(I) & = & N_0 \,\partial_x(I) + z\, N_0 \,\partial_y(I)\, + M_0 \,\partial_z(I) = 0,  \\ [2mm]
\chi(I) & = & f \,\partial_x(I) + g \,\partial_y(I)\, + h \,\partial_z(I) = 0. \label{dosi2}
\end{eqnarray}
where $M_0$ and $N_0$ are, respectively, the numerator and denominator of the 2ODE (\ref{hyp2ode}). Defining $D_1 \equiv f\,D_0 - N_0\,\chi$, and applying it to $I$ we obtain:
\begin{equation}
\label{d1i}
D_1 (I) = ( f\,D_0 - N_0\,\chi) (I) = (z\,f-g)\,N_0 \,\partial_y(I) + (M_0\,f - N_0\,h)\,\partial_z(I) = 0.
\end{equation}
From eq.(\ref{d1i}) we have (see \cite{Noscpc2019}) that $I_y/I_z=-(M_0\,f - N_0\,h)/(N_0\, (z\,f-g))=S_1. \,\,\,\Box$

\begin{obs}
We note that the function $S_1$ as presented in the equation {\em (\ref{s12ode1})} `defines', in a certain sense, the relationship between the vector field $\chi$ and the 2ODE {\em (\ref{hyp2ode})}. So, we can use it to produce an algorithm (semi) to find the 2ODE ($M_0$ and $N_0$) and, in view of {\em (\ref{s12ode1})}, the $S$-function.
\end{obs}

\bigskip

\subsection{A possible method}
\label{mffi}

In this section, we will use the equation (\ref{s12ode1}) in the process of building an algorithm to find the general solution of 1ODE (\ref{1odelem}). Before describing the steps of a possible algorithm in more detail, let's outline the main steps of the process and discuss its application to some examples in order to materialize (and clarify) the path we are trying to follow to achieve our goal.

\medskip

\noindent
{\bf The Procedure:} (Sketch)
\begin{itemize}
\item If the 1ODE can be placed in the form (\ref{1odelem}) then construct the operator $D$ (\ref{eqitem3}).
\item Make the substitution $\theta \rightarrow z$ on the $D$ operator and find the polynomial operator $\chi$ (\ref{vfchi}).
\item Construct polynomial candidates $M_c$ and $N_c$ (with undetermined coefficients $\{m_i\}$ and $\{n_j\}$) of some chosen degree in the variables $(x,y,z)$.
\item Substitute

\noindent
$S_1 = \displaystyle{\frac{M_c\,f - N_c\,h}{N_c\,(g-z\,f)}}$

\noindent
in the equation for the $S$-function:

\noindent
$D_x(S_1) = {S_1}^2+ \displaystyle{\frac{\partial \Phi_c}{\partial z}\,S_1 - \frac{\partial \Phi_c}{\partial y}},$

\noindent
where $D_x \equiv \partial_x+z\,\partial_y+\Phi_c\,\partial_z$ and $\Phi_c \equiv M_c/N_c$.
\item Solve the polynomial equation for the unknown coefficients $\{m_i\}$ and $\{n_j\}$, obtaining $S_1$.
\item Use the $S$-function method to find de Liouvillian first integral $I$ of the 2ODE $z' = \Phi(x,y,z)$.
\item Make the substitution $z \rightarrow \theta$ on the first integral $I$ and obtain the desired general solution of the 1ODE.
\end{itemize}

\medskip

Before presenting a more formal algorithm let's discuss some examples:

\noindent
{\bf Example 1:} Consider the 1ODE given by
\begin{equation}
\label{exem1}
\frac{dy}{dx} = {\frac {{{\rm e}^{x}}{x}^{3}{y}^{2}+{{\rm e}^{x}}{x}^{2}{y}^{2}+2\,{
{\rm e}^{x}}{x}^{2}y+{{\rm e}^{x}}xy+{{\rm e}^{x}}x+{y}^{2}+{{\rm e}^{
x}}}{{x}^{2}{y}^{2}+{{\rm e}^{x}}{x}^{2}+xy+1}}
\end{equation}
and let's apply the procedure sketched above:

\begin{itemize}

\item Choosing $\theta = {\rm e}^x$, we have that $\theta_x = {\rm e}^x = \theta$ and $\theta_y = 0$. So, the operator $D$ will be:
\begin{equation}
\label{ex1D}
D = \partial_x + {\frac {\theta\,{x}^{3}{y}^{2}+\theta\,{x}^{2}{y}^{2}+2\,\theta\,{x}^{2}y+\theta\,xy+\theta\,x+{y}^{2}+\theta}{{x}^{2}{y}^{2}+\theta\,{x}^{2}+xy+1}}\,\partial_y + \theta\,\partial_{\theta}.
\end{equation}

\item The polynomials $f$, $g$ and $h$ will be (see eq. (\ref{fghnovo})):
\begin{eqnarray}
\label{ex1f}
f &=& {x}^{2}{y}^{2}+z{x}^{2}+xy+1,
 \\ [2mm]
\label{ex1g}
 g &=& z{x}^{3}{y}^{2}+z{x}^{2}{y}^{2}+2\,z{x}^{2}y+zxy+zx+{y}^{2}+z,
 \\ [2mm]
\label{ex1h}
 h &=& z \left( {x}^{2}{y}^{2}+z{x}^{2}+xy+1 \right).
\end{eqnarray}

The operator $\chi$ will be:
\begin{equation}
\label{ex1chi}
\chi = f\, \partial_x + g\,\partial_y + h\,\partial_{z}.
\end{equation}

\item For the degrees ${\rm deg}_M=4$ and ${\rm deg}_N=5$ we obtain:
\begin{eqnarray}
\label{ex1M}
M_0 &=& -\left( xz-y \right)  \left( xz+y \right),
 \\ [2mm]
\label{ex1N}
 N_0 &=& - x \left( xy+1 \right) ^{2},
 \\ [2mm]
\label{ex1S1}
 S_1 &=& - \frac{{x}^{2}{y}^{2}+{x}^{2}z+xy+1}{x \left( xy+1 \right) ^{2}}
\end{eqnarray}

\item Using $S_1$ in the $S$-function method, we have the following first integral:
\begin{equation}
\label{ex1I}
I = {\rm e}^{\frac{1}{xy+1}} \left( zx-y \right)
\end{equation}

\item Making the substitution $z={\rm e}^x$ we finally arrive at the solution:
\begin{equation}
\label{ex1sol}
{\rm e}^{{\frac{1}{xy+1}}} \left( x\,{\rm e}^x -y \right) = C
\end{equation}

\end{itemize}

\noindent
Now let's take a closer look at the derivatives of the first integral $I$ (of the vector field $\chi$):
\begin{eqnarray}
I_x &=& \frac {{\rm e}^{\frac{1}{xy+1}}}{(xy+1)^2}\,\left( z{x}^{2}{y}^{2}+
zxy+{y}^{2}+z \right),
\label{ixex1} \\ [2mm]
I_y &=& - \frac {{\rm e}^{\frac{1}{xy+1}}}{(xy+1)^2} \, \left( {x}^{2}{y}^{2}+
z{x}^{2}+xy+1 \right),
\label{iyex1} \\ [2mm]
I_z &=& \frac {{\rm e}^{\frac{1}{xy+1}}}{(xy+1)^2} \, x (xy+1)^2.
\label{izex1}
\end{eqnarray}
From them we can determine (directly) the integrating factor $R$, the 2ODE associated ($\Phi$), the polynomials $P_1$, $P_2$ and $P_3$ and the $S$-functions $S_1$, $S_2$ and $S_3$:
\begin{eqnarray}
\Phi &=& {\frac {{x}^{2}{z}^{2}-{y}^{2}}{ \left( xy+1 \right) ^{2}x}},
\label{Phiex1} \\ [2mm]
R &=& \frac {{\rm e}^{\frac{1}{xy+1}}}{(xy+1)^2},
\label{rex1} \\ [2mm]
P_1 &=& {x}^{2}{y}^{2}z+xyz+{y}^{2}+z,
\label{p1ex1} \\ [2mm]
P_2 &=& - \left( {x}^{2}{y}^{2}+{x}^{2}z+xy+1 \right),
\label{p2ex1} \\ [2mm]
P_3 &=&  x (xy+1)^2,
\label{p3ex1} \\ [2mm]
S_1 &=& -{\frac {{x}^{2}{y}^{2}+z{x}^{2}+xy+1}{ \left( xy+1 \right) ^{2}x}},
\label{s1ex1} \\ [2mm]
S_2 &=&{\frac {z{x}^{2}{y}^{2}+zxy+{y}^{2}+z}{ \left( xy+1 \right)^{2}x}},
\label{s2ex1} \\ [2mm]
S_3 &=&  -{\frac {z{x}^{2}{y}^{2}+zxy+{y}^{2}+z}{{x}^{2}{y}^{2}+z{x}^{2}+xy+1}}.
\label{s3ex1}
\end{eqnarray}
The components $(f,g,h)$ of the vector field $\chi$ are (see above)
\begin{eqnarray}
f &=& {x}^{2}{y}^{2}+z{x}^{2}+xy+1,
 \\ [2mm]
g &=& z{x}^{3}{y}^{2}+z{x}^{2}{y}^{2}+2\,z{x}^{2}y+zxy+zx+{y}^{2}+z,
 \\ [2mm]
h &=& z \left( {x}^{2}{y}^{2}+z{x}^{2}+xy+1 \right).
\end{eqnarray}

\begin{obs}
\label{ex1obs}
So, comparing $(f,g,h)$ with the polynomials $P_1$, $P_2$ and $P_3$ and with the numerators and denominators of the $S$-functions, we can see that:
\begin{equation}
N_0 = P_3,\,\,\,\, f = - P_2,\,\,\,\, h = - z\,P_2.
\end{equation}
The first of these equations may not surprise, since $\Phi = -(P_1+z\,P_2)/P_3$. However, the following two equations can seem (at first glance) intriguing. More about that later (please see subsection \ref{api}).
\end{obs}

\medskip

\noindent
{\bf Example 2:} Consider now the 1ODE given by
\begin{equation}
\label{exem2}
\frac{dy}{dx} ={\frac {{-{\rm e}^{y}} \left( {{\rm e}^{2y}}{x}^{2}y-2\,{{\rm e}^{y}}x{y}^{2}-x{{\rm e}^{y}}y+{y}^{3}-1 \right) }
{ {{\rm e}^{3y}}{x}^{3}y\!+\! {{\rm e}^{3y}}{x}^{3}\!-2{{\rm e}^{2y}}{x}^{2}{y}^{2}\!-3{{\rm e}^{2y}}{x}^{2}y\!+{{\rm e}^{y}}x{y}^{3}\!
+{{\rm e}^{y}}x{y}^{2}\!+x{{\rm e}^{y}}y\!-x{{\rm e}^{y}}\!+\!1}}.
\end{equation}
Choosing $z = \theta = {\rm e}^y$ and applying the whole procedure again we obtain

\begin{eqnarray}
\label{ex2f}
f &=&\! {x}^{3}y{z}^{3}\!+\!{z}^{3}{x}^{3}\!-\!2{x}^{2}{y}^{2}{z}^{2}\!-\!3{z}^{2}{x}^{2}y\!+\!x{y}^{3}z\!+\!zx{y}^{2}\!+\!xzy\!-\!zx\!+\!1,
 \\ [2mm]
\label{ex2g}
 g &=& -z \left( {z}^{2}{x}^{2}y-2\,zx{y}^{2}-xzy+{y}^{3}-1 \right),
 \\ [2mm]
\label{ex2h}
 h &=& -{z}^{2} \left( {z}^{2}{x}^{2}y-2\,zx{y}^{2}-xzy+{y}^{3}-1 \right).
\end{eqnarray}

\noindent
The function $S_1$ is
\begin{equation}
\label{ex2S1}
 S_1 = {\frac {{z}^{3}{x}^{3}-2\,{z}^{2}{x}^{2}y+zx{y}^{2}+xzy+1}{x \left( {z
}^{2}{x}^{2}y-2\,zx{y}^{2}-xzy+{y}^{3}-1 \right) }}.
\end{equation}

\noindent
From $S_1$ we can determine the first integral (for the vector field $\chi$) and (making the substitution $z={\rm e}^y$) the general solution of the 1ODE (\ref{exem2}):
\begin{equation}
\label{ex2Isol}
I = {\rm e}^{\frac{1}{zx-y}} \left( xzy+1 \right), \,\,\,\, {\rm e}^{\frac{1}{{\rm e}^yx-y}} \left( x{{\rm e}^{y}}y+1 \right)=C.
\end{equation}

\medskip

\noindent
From the derivatives of the first integral $I$ we can get $\Phi$, the polynomials $P_1$, $P_2$ and $P_3$ and the $S$-functions $S_1$, $S_2$ and $S_3$:
\begin{eqnarray}
\Phi &=& {\frac { \left( zx-y \right) ^{2} \left( zx+y \right) z}{x \left( {z}^
{2}{x}^{2}y-2\,zx{y}^{2}-xzy+{y}^{3}-1 \right) }},
\label{Phiex2} \\ [2mm]
P_1 &=& z \left( {z}^{2}{x}^{2}y-2\,zx{y}^{2}-xzy+{y}^{3}-1 \right),
\label{p1ex1} \\ [2mm]
P_2 &=& {z}^{3}{x}^{3}-2\,{z}^{2}{x}^{2}y+zx{y}^{2}+xzy+1,
\label{p2ex1} \\ [2mm]
P_3 &=&  x \left( {z}^{2}{x}^{2}y-2\,zx{y}^{2}-xzy+{y}^{3}-1 \right),
\label{p3ex1} \\ [2mm]
S_1 &=& {\frac {{z}^{3}{x}^{3}-2\,{z}^{2}{x}^{2}y+zx{y}^{2}+xzy+1}{x \left( {z
}^{2}{x}^{2}y-2\,zx{y}^{2}-xzy+{y}^{3}-1 \right) }},
\label{s1ex1} \\ [2mm]
S_2 &=&{\frac {z}{x}},
\label{s2ex1} \\ [2mm]
S_3 &=&  {\frac {z \left( {z}^{2}{x}^{2}y-2\,zx{y}^{2}-xzy+{y}^{3}-1 \right) }{
{z}^{3}{x}^{3}-2\,{z}^{2}{x}^{2}y+zx{y}^{2}+xzy+1}}.
\label{s3ex1}
\end{eqnarray}

\medskip

\begin{obs}
\label{ex2obs}
Again, if we compare $(f,g,h)$ with the polynomials $P_1$, $P_2$ and $P_3$ and with the $S$-functions, we see that:
\begin{equation}
N_0 = P_3,\,\,\,\, g = - P_1,\,\,\,\, h = - z\,P_1.
\end{equation}
Besides that, as the roles of $x$ and $z$ can be switched in the first integral $I$ (i.e., the transformation $x \rightarrow z, \, z\rightarrow x$ is a symmetry transformation for $I$) the $S$-function $S_2$ has a very simple format.
\end{obs}

\medskip

{\bf Example 3:} Let's see now an example where we have a function of $x$ and $y$. Consider the 1ODE
\begin{equation}
\label{exem3}
\frac{dy}{dx} =\frac{xy\ln  \left( {\frac {x}{y}} \right) + \left( x{y}^{5}-2\,{x}^{2}{y}^{
3}+{x}^{3}y+{y}^{4}+{x}^{2}y-2\,x{y}^{2}+{x}^{2} \right) y}
{2\,x{y}^{2}\ln  \left( {\frac {x}{y}} \right) +x \left( -x{y}^{5}+2\,{
x}^{2}{y}^{3}-{x}^{3}y+2\,x{y}^{3}+{y}^{4}-2\,x{y}^{2}+{x}^{2} \right)}.
\end{equation}

Choosing $z = \theta = \ln  \left( {\frac {x}{y}} \right)$ and applying the whole procedure again we obtain

\begin{eqnarray}
\label{ex3f}
\!\!\!\!f\! &=&\!\! x \left( x{y}^{5}-2\,{x}^{2}{y}^{3}+{x}^{3}y-2\,x{y}^{3}-{y}^{4}+2\,x{y}^{2}-2\,z{y}^{2}-{x}^{2} \right),
 \\ [2mm]
\label{ex3g}
\!\!\!\! g\! &=&\! - \left( x{y}^{5}-2\,{x}^{2}{y}^{3}+{x}^{3}y+{y}^{4}+{x}^{2}y-2\,x{y}^{2}+{x}^{2}+zx \right) y,
 \\ [2mm]
\label{ex3h}
\!\!\!\! h\! &=&\! 2\,x{y}^{5}-4\,{x}^{2}{y}^{3}+2\,{x}^{3}y-2\,x{y}^{3}+{x}^{2}y-2\,z{y}^{2}+zx.
\end{eqnarray}

\noindent
The function $S_1$ is
\begin{equation}
\label{ex3S1}
 S_1 = {\frac {x{y}^{4}-2\,{x}^{2}{y}^{2}+{x}^{3}-2\,x{y}^{2}-2\,yz}{{y}^{4}-
2\,x{y}^{2}+{x}^{2}}}.
\end{equation}

\noindent
The first integral and the general solution of the 1ODE are
\begin{equation}
\label{ex3Isol}
I = {\rm e}^{\frac{1}{y^2-x}} \left( xy+z \right), \,\,\,\, {\rm e}^{\frac{1}{y^2-x}} \left( xy+\ln  \left( {\frac {x}{y}} \right) \right)=C.
\end{equation}

\noindent
The polynomials $P_1$, $P_2$, $P_3$ and $\Phi$ are:
\begin{eqnarray}
\!\!\!P_1\!\! &=&\!\! {y}^{5}-2\,x{y}^{3}+{x}^{2}y+xy+z,
\label{p1ex3} \\ [2mm]
\!\!\!P_2\!\! &=&\!\! x{y}^{4}-2\,{x}^{2}{y}^{2}+{x}^{3}-2\,x{y}^{2}-2\,yz,
\label{p2ex3} \\ [2mm]
\!\!\!P_3\!\! &=&\!\! \left( -{y}^{2}+x \right) ^{2},
\label{p3ex3} \\ [2mm]
\!\!\!\Phi\!\! &=&\!\! {\frac {x{y}^{4}z\!-\!2{x}^{2}{y}^{2}z\!+\!{y}^{5}\!+\!{x}^{3}z\!-\!2x{y}^{3}\!-\!2x
{y}^{2}z\!+\!{x}^{2}y\!-\!2y{z}^{2}\!+\!xy\!+\!z}{ \left( -{y}^{2}+x \right) ^{2}}}.
\label{Phiex3}
\end{eqnarray}

\begin{obs}
\label{ex3obs}
Some comments:
\begin{enumerate}
\item This time, looking at the polynomials $P_1$, $P_2$ and $P_3$ in the example 3, we see that none of them is a member of $\{c\,f,c\,g,c\,h\}$ where $c$ is a constant.

\item A very noticeable difference between examples 1 and 2 and example 3 is that in examples 1 and 2 the elementary function present on the 1ODE was a function of only one variable ($x$ or $y$) whereas in example 3 the function $\theta$ was a function of the two variables $(x,y)$, i.e., $\theta=\theta(x,y)$.

\item In general it is much simpler (computationally speaking) to calculate the $S$-function if we already know its numerator or denominator. In these cases the algorithm is much more efficient (i.e., faster and with less memory consumption).

\item What happened in examples 1 and 2 was not a fluke. We can show that {\bf (please see subsection \ref{api})} below, 
, for a class of 1ODEs (that we will cal $L_S$), if $\theta$ is an elementary function of only one variable, then one of the coefficients of the vector field $\chi$ will divide one of the polynomials $(P_1,P_2,P_3)$.

\item The good news is that, if $\theta$ is an elementary function of a rational function, 
for 1ODE $\in L_S$ , we can perform a variable transformation that takes $\theta(x,y)$ into $\overline{\theta}(x)$ (or $\overline{\theta}(y)$).
\end{enumerate}
\end{obs}

\subsection{A possible improvement}
\label{api}

What was stated in observation 4 (remark \ref{ex3obs}) can be demonstrated and (in conjunction with the fact presented in observation 5) can be used to greatly improving the efficiency of the algorithm in a wide variety of cases. Before presenting the alternative method, let's define (a little more) the type of 1ODE that these improvement can handle.

\begin{defin}
\label{lsset}
Consider that an 1ODE as described by equation {\em (\ref{1odelem})} has the following characteristics:
\begin{enumerate}
\item  $I(x,y,\theta(x,y)) = c$ ($c$ constant) represents a general solution of the 1ODE {\em (\ref{1odelem})} and $I$ is a Liouvillian function of $(x,y,\theta)$.
\item The function $\theta(x,y)$ is in an elementary extension $E$ of the differential field (of characteristic zero) $\C(x,y)$ such that $E=\C(x,y,\exp(r))$ or $E=\C(x,y,\ln(r))$, where $r \in \C(x,y)$.
\item The 1ODE is associated with a vector field $\chi \equiv f\,\partial_x+g\,\partial_y+h\,\partial_z$ such that $f,g,h \in \C[x,y,z]$ and
\begin{eqnarray}
f &=& l_d |_{\theta \rightarrow z}, \nonumber
\\ [2mm]
g &=& l_d\,\phi |_{\theta \rightarrow z}, \nonumber
\\ [2mm]
h &=& l_d\,\left(\theta_x + \phi\,\theta_y\right) |_{\theta \rightarrow z}. \nonumber
\end{eqnarray}
\item The derivatives of the first integral $I$ (of the vector field $\chi$) are of the form: $\partial_k (I) = \exp(A/B)\,\prod_i p_i^{n_i}\,P_k,\,k \in \{1,2,3\}$, where $A,B,p_i,P_k$ are polynomials in $\C[x,y,z]$ and $n_i$ are integers.
\item Let $T$ be the transformation $T=\{u=r(x,y),v=y\}$ (or $T = \{u = x, v = r(x,y) \}$), where $r \in \C(x,y)$ is the argument of $\theta$. If $T^{-1} = \{x=r_1(u,v),y=v\}$ (or $T^{-1} = \{ x = u, y=r_1(u,v) \}$) denote its inverse, we are assuming that $r_1$ is a rational function of $(u,v)$, i.e., $r_1 \in \C(u,v)$.
\end{enumerate}
We will denote the set of 1ODEs that have these characteristics as $L_S$.
\end{defin}

Now, we are going to present a result that, if applied to an 1ODE $\in L_S$, can be used to construct an improvement of the method sketched above.

\begin{teor}
\label{theoAimp}
Let $\chi \equiv f\,\partial_x+g\,\partial_y+h\,\partial_z$ be the polynomial vector field associated with the 1ODE {\em (\ref{1odelem})} ($\phi = M/N,\,M,N \in   \C[x,y,\theta]$) and let the first integral $I$ (of the vector field $\chi$) be such that its derivatives are of the form: $\partial_k (I) = \exp(A/B)\,\prod_i p_i^{n_i}\,P_k,\,k \in \{1,2,3\}$, where $A,B,p_i,P_k$ are polynomials in $\C[x,y,z]$ and $n_i$ are rational numbers. We can state that:

\noindent
i) If $\,\theta = {\rm e}^x\,$ then $\,f | P_2,\,$ $h=z\,f\,$  and $\,g |(P_1 + z\,P_3)$.

\noindent
ii) If $\,\theta = \ln(x)\,$ then $\,h | P_2,\,$ $f=x\,h\,$ and $\,g | (x\,P_1+P_3)$.

\noindent
iii) If $\,\theta = {\rm e}^y\,$ then $\,g | P_1,\,$ $h=z\,g\,$ and $\,f |(P_2 + z\,P_3)$.

\noindent
iv) If $\,\theta = \ln(y)\,$ then $\,h | P_1,\,$ $g=y\,h\,$ and $\,f | (y\,P_2+P_3)$.
\end{teor}

\noindent
{\it Proof.} We have seen that we can write the operator $D \equiv \partial_x + \phi\,\partial_y$ -- associated with the 1ODE (\ref{1odelem}) -- in the form
$
D ={\partial_x} +  \phi\,{\partial_y} + \left(\theta_x + \phi\,\theta_y\right)\,{\partial_\theta},
$
where $\theta_x$ and $\theta_y$ are rational functions of $(x,y,\theta)$.
In order to obtain the vector field $\chi$ we have to multiply $D$ by the ${\rm lcm}$ of the denominators of $\phi$, $\theta_x$ and $\theta_y\,\phi$ and make the substitution $\theta \rightarrow z$, obtaining
$
\chi = f\,\partial_x + g\,\partial_y + h\,\partial_z, \nonumber
$
where $f,\,g$ and $h$ are polynomials given by
\begin{eqnarray}
f &=& l_d |_{\theta \rightarrow z}, \nonumber
\\ [2mm]
g &=& l_d\,\phi |_{\theta \rightarrow z}, \nonumber
\\ [2mm]
h &=& l_d\,\left(\theta_x + \phi\,\theta_y\right) |_{\theta \rightarrow z}. \nonumber
\end{eqnarray}

\noindent
Now we can prove the statements:

\medskip

\noindent
$i$): If $\theta={\rm e}^x$ then $\theta_x = z$ and $\theta_y = 0$ and so
\begin{equation}
\label{proo1}
f=N(x,y,z),\,\,g=M(x,y,z), \,\,h=z\,N(x,y,z).
\end{equation}
We have that $\chi(I)\! =\! f\,I_x+g\,I_y+h\,I_z= fRP_1+gRP_2+hRP_3 \,{\rm (by\, hypothesis)} = R\, (f\,P_1+g\,P_2+h\,P_3) = 0$. Substituting (\ref{proo1}) in $\chi(I)$ and noting that $R$ is not null we can write $N\,P_1+M\,P_2+z\,N\,P_3=0$ implying that $N\,(P_1+z\,P_3)=-M\,P_2$. Therefore, we can write
\begin{equation}
\label{proo2}
P_1+z\,P_3=- \frac{M\,P_2}{N} \,\, {\rm and} \,\, P_2=- \frac{N\,(P_1+z\,P_3)}{M}.
\end{equation}
Since $M$ and $N$ are coprime, we have that $\,N|P_2\,\,\Rightarrow\,\,f | P_2\,$ and $\,M|(P_1 + z\,P_3)\,\,\Rightarrow\,\,g |(P_1 + z\,P_3)$.

\medskip

\noindent
$ii$): If $\theta=\ln(x)\,\,\Rightarrow\,\,\theta_x = 1/x$ and $\theta_y = 0$. We have two cases:
\begin{equation}
\label{proo3}
f=x\,N(x,y,z),\,\,g=x\,M(x,y,z), \,\,h=N(x,y,z),
\end{equation}
if $x$ is not a factor of $N$ or
\begin{equation}
\label{proo4}
f=x^n\,P_N(x,y,z),\,\,g=M(x,y,z), \,\,h=x^{n-1}P_N(x,y,z),
\end{equation}
where $P_N=N/x^n$ is a polynomial, $n$ is a positive integer and $x$ and $P_N$ are coprime.

\noindent
First case ($x$ and $N$ are coprime): $N\,(x\,P_1+P_3)=-x\,M\,P_2$. Therefore, $\,N|P_2\,\,\Rightarrow\,\,h | P_2\,$ and $\,x\,M|(x\,P_1+P_3)\,\,\Rightarrow\,\,g |(x\,P_1+P_3)$.

\noindent
Second case ($x|N$): $x^{n-1}P_N\,(x\,P_1+P_3)=-\,M\,P_2$. Since $M$ and $N$ are coprime, we have that $x^{n-1}P_N\,(=h)$ and $M$ are coprime. Therefore, $h | P_2\,$ and $g |(x\,P_1+P_3)$.

\medskip

\noindent
$iii$): If $\theta={\rm e}^y$ then $\theta_x = 0$ and $\theta_y = z$ and so
\begin{equation}
\label{proo5}
f=N(x,y,z),\,\,g=M(x,y,z), \,\,h=z\,M(x,y,z).
\end{equation}
We can write $N\,P_1+M\,P_2+z\,M\,P_3=0$ implying that $N\,P_1=-M\,(P_2+z\,P_3)$. Therefore, we can write
\begin{equation}
\label{proo6}
P_2+z\,P_3=- \frac{N\,P_1}{M} \,\, {\rm and} \,\, P_1=- \frac{M\,(P_2+z\,P_3)}{N}.
\end{equation}
Since $M$ and $N$ are coprime, we have that $\,N|(P_2+z\,P_3)\,\,\Rightarrow\,\,f | (P_2+z\,P_3)\,$ and $\,M|P_1 \,\,\Rightarrow\,\,g | P_1$.

\medskip

\noindent
$iv$): If $\theta=\ln(y)\,\,\Rightarrow\,\,\theta_x = 0$ and $\theta_y = 1/y$. We have two cases:
\begin{equation}
\label{proo7}
f=y\,N(x,y,z),\,\,g=y\,M(x,y,z), \,\,h=M(x,y,z),
\end{equation}
if $y$ is not a factor of $M$ or
\begin{equation}
\label{proo8}
f=N(x,y,z),\,\,g=y^{n}P_M(x,y,z), \,\,h=y^{n-1}P_M(x,y,z),
\end{equation}
where $P_M=M/y^n$ is a polynomial, $n$ is a positive integer and $y$ and $P_M$ are coprime.

\noindent
{\bf First case;} ($y$ and $M$ are coprime): $N\,y\,P_1=-M\,(y\,P_2+P_3)$. Therefore, $\,M|P_1\,\,\Rightarrow\,\,h | P_1\,$ and $\,y\,N|(y\,P_2+P_3)\,\,\Rightarrow\,\,f |(y\,P_2+P_3)$.

\medskip
\noindent
{\bf Second case:} ($y|M$): $y^{n-1}P_M\,(y\,P_2+P_3)=-\,N\,P_1$. Since $M$ and $N$ are coprime, we have that $y^{n-1}P_M\,(=h)$ and $N$ are coprime. Therefore, $h | P_1\,$ and $f |(y\,P_2+P_3)$. $\,\,\,\Box$

\medskip

\begin{cor}
\label{coroAimp}
If the 1ODE {\em (\ref{1odelem})} $\in L_S$, then there is a rational transformation $T$ (with rational inverse $T^{-1}$) such that the transformed 1ODE$_{[tr]}$ has an associated vector field $\chi_{[tr]}$ in which $f_i|P_j$ for some $\,f_i \in \{f,g,h\}\,$ and for some $\,P_j \in \{P_1,P_2\}$.
\end{cor}

\noindent
{\it Proof.} For any 1ODE $\in L_S$ we can perform a variable transformation that takes $\theta(x,y)$ into $\exp(x)$, $\ln(x)$, $\exp(y)$ or $\ln(y)$. Thus, any 1ODE $\in L_S$ can be transformed into an 1ODE that fulfills the premises of the theorem \ref{theoAimp}. Therefore, the conclusion is a direct consequence of the theorem \ref{theoAimp}. $\,\,\,\Box$

\medskip

In this point we can use the compatibility conditions for the first integral $I$ of the vector field $\chi$ to write the 1PDE for $S$ in terms of $f,\, g$ and $h$.

\begin{teor}
\label{theoSsis}
Let $\chi \equiv f\,\partial_x+g\,\partial_y+h\,\partial_z$ be the polynomial vector field associated with the 1ODE {\em (\ref{1odelem})} and let $I$ be a first integral it. Then the $S$-function associated with $\chi$ through $I$ obeys the 1PDE given by
\begin{equation}
\label{eqSsys}
\chi (S)=S^2\left(\frac{fg_z-gf_z}{f}\right)+S\left(\frac{gf_y-fg_y+fh_z-hf_z}{f}\right)-\left(\frac{fh_y-hf_y}{f}\right).
\end{equation}
\end{teor}

\noindent
{\it Proof.} See the proof of theorem 1.1 and corollary 1.1 of \cite{Noscpc2019}. $\,\,\,\Box$

\medskip

\noindent
Since $S=I_y/I_z=P_2/P_3$ then, if a 1ODE $\in L_S$ we can perform a simple variable transformation to turn $\theta$ into a function of just one variable, in which case we know one of the polynomials that form the function $S$ (see theorem \ref{theoAimp}) and, in this way, we can look for only one polynomial instead of two in the equation (\ref{eqSsys}). Below we will present a series of steps that can be performed for every 1ODE $\in L_S$. The first of these transformations aims to transform the function $\theta$ into a function of only one variable. The second transformation is just to standardize the algorithm in the sense of making it work only with the function $\ln(x)$. Finally, the third transformation lends itself only to making the known polynomial the denominator of the function $S$, that is, $P_3$.

\begin{itemize}

\item In first place we can use the rational transformation $T_1$ such that its inverse is given by ${T_1}^{-1}=\{y_1=x,x_1=r(x,y)\}$ (where $r(x,y)$ is the argument of $\theta$) to obtain an 1ODE$_{[1]}$ such that $\theta_{[1]} = \exp(x_1)$ or $\theta_{[1]} = \ln(x_1)$.

\item If $\theta_{[1]} = \exp(x_1)$ we can use the transformation $T_2=\{x_1=\ln(x_2),y_1=y_2\}$ and obtain an 1ODE$_{[2]}$ such that $\theta_{[2]} = \ln(x_2)$. If $\theta_{[1]} = \ln(x_1)$ we can use the transformation $T_2=\{x_1=(x_2,y_1=y_2\}$ and obtain an 1ODE$_{[2]}$ such that $\theta_{[2]} = \ln(x_2)$.

\begin{obs}
\label{transft2}
The transformation $T_2$ serves only to standardize (computationally speaking) the procedure, since, in this way, we will deal only with the function $\ln(x)$.
\end{obs}

\item For the 1ODE$_{[2]}$, the vector field $\chi_2$ associated is given by $f_2\,\partial_{x_2}+g_2\,\partial_{y_2}+h_2\,\partial_{z_2}$, where (To make the notation less heavy we omitted index 2 in the following equations):
\begin{equation}
f=x\,N(x,y,z),\,\,g=x\,M(x,y,z), \,\,h=N(x,y,z),
\end{equation}
if $x$ is not a factor of $N$ or
\begin{equation}
f=x^n\,P_N(x,y,z),\,\,g=M(x,y,z), \,\,h=x^{n-1}P_N(x,y,z),
\end{equation}
where $P_N=N/x^n$ is a polynomial, $n$ is a positive integer and $x$ and $P_N$ are coprime (see case ($ii$) of theorem \ref{theoAimp}).

\item Since the status of the variables $(x_2,y_2,z_2)$ in the vector field $\chi_2$ is the same, we can make the transformation $T_3=\{x_2=y_3,y_2=z_3,z_2=x_3\}$ that will make $P_2$ take the place of $P_3$ (which is the denominator of the transformed $S$-function). The transformed vector field $\chi_3$ will be given by $f_3\,\partial_{x_3}+g_3\,\partial_{y_3}+h_3\,\partial_{z_3}$, where
\begin{eqnarray}
f_3 &=& T_3(h_2), \nonumber
\\ [2mm]
g_3 &=& T_3(f_2), \nonumber
\\ [2mm]
h_3 &=& T_3(g_2). \nonumber
\end{eqnarray}

\begin{obs}
\label{transft3}
The transformation $T_3$ only aims to make the known polynomial be the denominator of the $S$-function.
\end{obs}

\item At this point we can use the equation (\ref{eqSsys}) to determine the numerator of the transformed $S$-function $S_{1\,[3]}$. We can use $S_{1\,[3]}$ to find a first integral $I_{[3]}$ for the vector field $\chi_{[3]}$.

\item Applying the inverse trasformation ${T_3}^{-1}$ to $I_{[3]}$ we obtain the first integral $I_{[2]}$ for $\chi_{[2]}$.  From $I_{[2]}$ we can write the general solution of 1ODE$_{[2]}$ (using $z\,\rightarrow\,\ln(x)$).

\item Finally, applying the inverse trasformation ${T_2}^{-1}$ followed by ${T_1}^{-1}$ to $I_{[2]}=c$, we will obtain $I = {T_1}^{-1}({T_2}^{-1}(I_{[2]}))=c$, which represents the general solution of the original 1ODE.

\end{itemize}

\noindent
Let's make these steps clearer with an example:

\bigskip

\noindent
{\bf Example 4:} Consider the 1ODE given by
\begin{equation}
\label{exem4}
\frac{dy}{dx} = {\frac {2\, \left( {{\rm e}^{xy}} \right) ^{2}{x}^{3}{y}^{2}-{{\rm e}^
{xy}}{x}^{3}{y}^{2}+{{\rm e}^{xy}}x{y}^{3}-5\,{{\rm e}^{xy}}{x}^{2}y+{
{\rm e}^{xy}}{y}^{2}+2\,x}{ \left( {{\rm e}^{xy}} \right) ^{2}{x}^{2}{
y}^{2}+{{\rm e}^{xy}}{x}^{4}y-{{\rm e}^{xy}}{x}^{2}{y}^{2}+{{\rm e}^{x
y}}{x}^{3}-3\,xy{{\rm e}^{xy}}+1}}.
\end{equation}

\begin{itemize}

\item In first place, let's make the transformation $T_1 = \{x=x/y,\,y=y\}$ leading to:
\begin{equation}
\label{1ode1ex4}
\!\!\!\!\frac{dy}{dx} \!=\! {\frac {y \left( 2\,{{\rm e}^{2\,x}}{x}^{3}+{{\rm e}^{x}}x{y}^{3}-{
{\rm e}^{x}}{x}^{3}+{{\rm e}^{x}}{y}^{3}-5\,{{\rm e}^{x}}{x}^{2}+2\,x
 \right) }{{{\rm e}^{2\,x}}{x}^{2}{y}^{3}+2\,{{\rm e}^{2\,x}}{x}^{4}-2
\,{{\rm e}^{x}}x{y}^{3}-4\,{{\rm e}^{x}}{x}^{3}+{y}^{3}+2\,{x}^{2}}},
\end{equation}
where now $\theta$ is a function of just one variable.

\item Next we're going to apply the transformation $T_2 = \{x= \ln(x),\,y=y\}$ obtaining:
\begin{equation}
\label{1ode2ex4}
\!\!\!\!\!\!\!\!\frac{dy}{dx} \!=\! {\frac {y ( 2\,{x}^{2} \ln(x) ^{3}
+x\ln(x){y}^{3}-x \ln(x) ^{3}+x{y}^{3}-5\,x \ln(x) ^{2}
+2\,\ln(x) ) }{ ( {x}^{2} \ln(x) ^{2}{y}^{3}\!+\!2\,{x}^{2} \ln(x) ^{4}\!-\!2\,x\ln(x){y}^{3}\!-\!4\,x \ln(x) ^{3}+{y}^{3}+2\, \ln(x) ^{2} ) x}},
\end{equation}
where now $\theta = \ln(x)$.

\item The vector field $\chi_{[2]}$ associated with the 1ODE$_{[2]}$ is defined by:
\begin{eqnarray}
\label{ex4f2}
f_2 &=& \left( {y}^{3}+2\,{z}^{2} \right)  \left( xz-1 \right) ^{2}x,
 \\ [2mm]
\label{ex4g2}
g_2 &=& y \left( 2\,{x}^{2}{z}^{3}+zx{y}^{3}+x{y}^{3}-x{z}^{3}-5\,{z}^{2}x+2\,z \right),
 \\ [2mm]
\label{ex4h2}
h_2 &=& \left( {y}^{3}+2\,{z}^{2} \right)  \left( xz-1 \right) ^{2}.
\end{eqnarray}

\item Let's apply the transformation $T_3 = \{x = y, y = z, z = x\}$ obtaining:
\begin{eqnarray}
\label{ex4f3}
f_3 &=& \left( {z}^{3}+2\,{x}^{2} \right)  \left( xy-1 \right) ^{2},
 \\ [2mm]
\label{ex4g3}
g_3 &=& \left( {z}^{3}+2\,{x}^{2} \right)  \left( xy-1 \right) ^{2}y,
 \\ [2mm]
\label{ex4h3}
h_3 &=& z \left( 2\,{x}^{3}{y}^{2}+xy{z}^{3}-{x}^{3}y+y{z}^{3}-5\,{x}^{2}y+2\,x \right).
\end{eqnarray}

\item Constructing a polynomial $P$ with arbitrary coefficients in $(x, y, z)$ and substituting in the equation (\ref{eqSsys}), we obtain (for degree 5):
\begin{equation}
\label{p23}
P = x\,z \left( -{z}^{3}+{x}^{2} \right)  \,\,\,\, \rightarrow \,\,\,\, S_{1\,[3]} = \frac{x\,z \left( -{z}^{3}+{x}^{2} \right)}{\left( {z}^{3}+2\,{x}^{2} \right)  \left( xy-1 \right) ^{2}},
\end{equation}
where, as we can see, it was only necessary to determine the polynomial $x\,z \left( -{z}^{3}+{x}^{2} \right)$, since the denominator is the (known) polynomial $f_3$.

\item Using the $S$-function method we can find a first integral for $\chi_{[3]}$:
\begin{equation}
\label{ii3}
I_{[3]} = -\ln  \left( {z}^{3}-{x}^{2} \right) +2\,\ln  \left( z \right) -
 \frac{1}{xy-1}.
\end{equation}

\item Applying ${T_3}^{-1}$ to $I_{[3]}$ we obtain:
\begin{equation}
\label{ii2}
I_{[2]} = -\ln  \left( {y}^{3}-{z}^{2} \right) +2\,\ln  \left( y \right) -
 \frac{1}{xz-1}.
\end{equation}

\item From $I_{[2]}$ we can write the general solution of 1ODE$_{[2]}$ ($z\,\rightarrow\,\ln(x)$):
\begin{equation}
\label{sol1ode2ex4}
-\ln  \left( {y}^{3}- \left( \ln  \left( x \right)  \right) ^{2}
 \right) +2\,\ln  \left( y \right) - \frac{1}{x\ln  \left( x \right) -1}=C_2.
\end{equation}

\item Applying ${T_2}^{-1}=\{x= {\rm e}^x,\,y=y\}$ to the general solution of 1ODE$_{[2]}$ we arrive at the general solution of 1ODE$_{[1]}$:
\begin{equation}
\label{sol1ode1ex4}
-\ln  \left( {y}^{3}-{x}^{2} \right) +2\,\ln  \left( y \right) -
 \frac{1}{x{{\rm e}^{x}}-1} = C_1.
\end{equation}

\item Finally, applying ${T_1}^{-1}=\{x= x\,y,\,y=y\}$ to the general solution of 1ODE$_{[1]}$ we have the general solution of the original 1ODE:
\begin{equation}
\label{sol1odeoex4}
-\ln  \left( -{x}^{2}{y}^{2}+{y}^{3} \right) +2\,\ln  \left( y
 \right) - \frac{1}{xy{{\rm e}^{xy}}-1} = C.
\end{equation}

\end{itemize}

\begin{obs}
Although the application of a sequence of transformations followed by the sequence of inverse transformations (applied in reverse order) could look a bit tiring to follow, in practice these steps have no algorithmic weight, i.e., they consume virtually no memory or CPU time. The important point is that the first transformation leads to a vector field whose associated $S$-function (to be determined) depends on obtaining only one polynomial (instead of 2). This greatly simplifies the algorithm.
\end{obs}

\subsection{Two possible algorithms}
\label{tpalg}

In this section we will present two possible algorithms to solve an 1ODE that presents elementary functions. The first (already briefly sketched in section \ref{mffi}) is generic. The second presents some restrictions (please see algorithm \ref{fastls} below) but is computationally incredibly efficient.

\begin{algor}[$Lsolv$] This algorithm is based on the method described in the section \ref{mffi}.
\label{lsolv}

\
\vspace{3mm}

{\bf Steps:}
\begin{enumerate}

\item Construct the vector field $\chi$, i.e., determine $[f,g,h]$ (by using definition \ref{lsset}).

\item Choose $DegMNP$ (a list of three positive integers - [$d_M$,$d_N$,$d_P$]) for the degrees of the candidates $M_c$, $N_c$ and $P_c$ (candidates for the polynomials $M_0$, $N_0$ and $P_3$).

\item Construct three polynomials $M_c$, $N_c$ and $P_c$ of degrees $d_M$, $d_N$ and $d_P$, respectively, with undetermined coefficients.

\item Substitute them in the equation $E_1\!: M_c\,f - N_c\,h+(z\,f-g)\,P_c=0$.

\item  Collect the equation $E_1$ in the variables $(x,y,z)$ obtaining a set of (linear) equations $S_{E_1}$ for the coefficients of the polynomials $M_c$, $N_c$ and $P_c$.

\item Solve $S_{E_1}$ to the undetermined coefficients.

\item Substitute the solution of $S_{E_1}\!\!$ in the equation $E_2\!:\! D_x(S_c) - {S_c}^2 - \partial_z( \Phi_c)\,S_c + \partial_y( \Phi_c)\!=\!0$, where $D_x \equiv \partial_x+z\,\partial_y+\Phi_c\,\partial_z$, $\Phi_c = M_c/N_c$ and $S_c = P_c/N_c$.

\item Collect the numerator of $E_2$ in the variables $(x,y,z)$ obtaining a set of equations $S_{E_2}$ for the remaining undetermined coefficients of the polynomials $M_c$, $N_c$ and $P_c$.

\item Solve $S_{E_2}$ for the remaining undetermined coefficients. If no solution is found then FAIL.

\item Substitute the solution of $S_{E_2}$ in $M_c$ and $N_c$ to obtain $M_0$, $N_0$ and in $P_c/N_c$ to obtain $S_1$.

\item Use the $S$-function method to find the Liouvillian first integral $I$ of the 2ODE $z' = \Phi(x,y,z)$.

\item Make the substitution $z \rightarrow \theta$ on the first integral $I$ and obtain the desired general solution of the 1ODE.

\end{enumerate}

\end{algor}

\medskip

\begin{obs}
Some comments:

\begin{itemize}

\item Since there is no bound for the degrees $[d_M,d_N,d_P]$ the procedure may never get to a conclusion, i.e., more formally, $Lsolv$ is a semi algorithm.


\end{itemize}

\end{obs}

\bigskip

\begin{algor}[$FastLs$]
\label{fastls}
This algorithm is an improvement of the algorithm \ref{lsolv} and is based on the theorems \ref{theoAimp} and \ref{theoSsis}, on the corollary \ref{coroAimp} and on their consequences (see example 4). It is restricted to 1ODEs $\in L_S$.
\
\vspace{3mm}

{\bf Steps:}
\begin{enumerate}

\item Use a variable transformation $T_1=\{x=r(x,y),y=y\}$ where $r(x,y)$ is the argument of $\theta$ to obtain an 1ODE$_{[1]}$ such that $\theta_{[1]} = \exp(x)$ or $\theta_{[1]} = \ln(x)$.

\item If $\theta_{[1]} = \exp(x)$ then make a transformation $T_2=\{x=\ln(x),y=y\}$ to obtain an 1ODE$_{[2]}$ such that $\theta_{[2]} = \ln(x)$.

\item Construct the vector field $\chi_2=f_2\,\partial_{x}+g_2\,\partial_{y}+h_2\,\partial_{z}$ associated with the 1ODE$_{[2]}$, where
\begin{equation}
f_2=x\,N(x,y,z),\,\,g_2=x\,M(x,y,z), \,\,h_2=N(x,y,z),
\end{equation}
if $x$ is not a factor of $N$ or
\begin{equation}
f_2=N(x,y,z),\,\,g_2=M(x,y,z), \,\,h_2=\frac{N(x,y,z)}{x},
\end{equation}
if $x|N$.

\item Apply the transformation $T_3 = \{x = y, y = z, z = x\}$ to the vector field $\chi_{[2]}$, obtaining $\chi_3 = f_3\,\partial_{x}+g_3\,\partial_{y}+h_3\,\partial_{z}$, where
\begin{eqnarray}
f_3 &=& T_3(h_2), \nonumber
\\ [2mm]
g_3 &=& T_3(f_2), \nonumber
\\ [2mm]
h_3 &=& T_3(g_2). \nonumber
\end{eqnarray}

\item Choose $d_P$ for the degree of the candidate $P$ (candidate for the polynomial $P_2$ -- numerator of the $S$-function).

\item Construct a polynomial $P$ in $(x, y, z)$ with arbitrary coefficients and substitute it in the equation $E_1\!: $ (\ref{eqSsys}).

\item  Collect the equation $E_1$ in the variables $(x,y,z)$ obtaining a set of quadratic equations $S_{E_1}$ for the coefficients of the polynomial $P$.

\item Solve $S_{E_1}$ to the undetermined coefficients. If no solution is found then FAIL.

\item Substitute the solution of $S_{E_1}\!\!$ in $P/f$ to obtain $S_{1\,[3]}$.

\item Use the $S$-function method to determine the first integral $I_{[3]}$ of the vector field $\chi_{[3]}$.

\item Apply the inverse trasformation ${T_3}^{-1}$ and obtain the first integral $I_{[2]}$ of the vector field $\chi_{[2]}$. From $I_{[2]}$ we can have (doing $z \rightarrow \ln(x)$) the general solution of the 1ODE$_{[2]}$.

\item Aplly ${T_2}^{-1}$ to the general solution of the 1ODE$_{[2]}$ and obtain the general solution of the 1ODE$_{[1]}$.

\item Aplly ${T_1}^{-1}$ to the general solution of the 1ODE$_{[1]}$ and obtain the general solution of the original 1ODE.
\end{enumerate}

\end{algor}

\medskip

\begin{obs}
The algorithm $FastLs$ uses the equation (\ref{eqSsys}) to determine the polynomial $P$ (the numerator of the $S$-function). Since (\ref{eqSsys}) has a term of the form $\,S^2$, we have that it can be written as (see the example 4):
\begin{equation}
\label{eqSsys2}
\chi \left(\frac{P}{f}\right)=\left(\frac{P}{f}\right)^2 \frac{fg_z-gf_z}{f}+\left(\frac{P}{f}\right) \frac{gf_y-fg_y+fh_z-hf_z}{f} - \frac{fh_y-hf_y}{f}.
\end{equation}
We can isolate the equation (\ref{eqSsys2}) for $P^2$, obtaining $\,P^2={\rm polynomial}$. Since the higher degree monomials of $P$ squared cannot cancel, the maximum degree of the polynomial $P$ is bounded. In this case, therefore, the part of $FastLs$ that determines the $S$-function is a full algorithm.
\end{obs}

\begin{obs}
The fact that $\,h | P_2,\,$ $f=x\,h\,$ and $\,g | (x\,P_1+P_3)$ (see theorem \ref{theoAimp}, case ($ii$)) does not imply that $f =c\, P_2$ and that $g = -c\,(P_1+z\,P_3)$ ($c$ constant). If the polynomials $P_2$ and $(x\,P_1+P_3)$ have a polynomial factor in common, the $FastLs$ algorithm may lose this case.
\end{obs}

\bigskip




\section{Performance}
\label{performance}
\hspace\parindent
%
The analysis presented in this section is divided in two parts\footnote{In this paper all the
computational data (time of running etc) was obtained on the same computer with the following configuration: Intel(R) Core(TM) i5-8265U @ 1.8 GHz.}:
\begin{itemize}
\item In the first subsection, we solve a set of 1ODEs 
that are very hard to solve using the traditional methods (and even non-traditional methods) implemented in computer algebra systems.
\item Finally, we present an example of a 1ODE with parameters. In that case we use the integrability analysis provided by our approach.
\end{itemize}

\subsection{A set of `hard' 1ODEs}
\label{hard1odes}
\hspace\parindent
In this section we present a set of 1ODEs which are hard to solve by other means (that not our method hereby introduced). In due time, just below the solution to the particular 1ODE presented, we will also show the CPU time and memory that it took to be solved using our methods with the hardware configuration already mentioned.
\medskip

Consider the following 1ODEs:
\begin{enumerate}

\item  The following five 1ODEs have an elementary Liouvillian solution. We will refer to them as 1ODEs 1, 2, 3, 4, 5:
\begin{equation}
\label{harde1}
\frac{dy}{dx}={\frac {x \left( 2\, \left( \ln  \left( x \right)  \right) ^{2}{x}^{4}
{y}^{2}+2\,\ln  \left( x \right) {x}^{2}y+2\,\ln  \left( x \right) {y}
^{2}-y{x}^{2}+{y}^{2}+2 \right) }{ \left( \ln  \left( x \right)
 \right) ^{2}{x}^{4}{y}^{2}+\ln  \left( x \right) {x}^{4}+\ln  \left(
x \right) {x}^{2}y+1}}
\end{equation}

\begin{equation}
\label{harde2}
\frac{dy}{dx}={\frac {3 \left( {{\rm e}^{x}} \right) ^{2}\!{x}^{4}\!{y}^{4}\!+\! \left( {y}^{4}\!-\!{
y}^{3}\!{x}^{4}\!-\!{x}^{3}\!{y}^{3}\!+\!6\,{x}^{3}\!{y}^{2}+x{y}^{4}\!
 \right) {{\rm e}^{x}}+3\,{x}^{2}}{ \left( {{\rm e}^{x}} \right) ^{2}{
x}^{2}{y}^{4}+ \left( {x}^{4}{y}^{2}-x{y}^{3}+2\,x{y}^{2} \right) {
{\rm e}^{x}}-{x}^{3}+y+1}}
\end{equation}

\begin{equation}
\label{harde3}
\frac{dy}{dx}=-{\frac { \left( 2\,{x}^{3}{y}^{4}-4\,{x}^{2}{y}^{2}-{x}^{2}+\ln
 \left( y \right) +2\,x \right) y}{-2\,{x}^{4}{y}^{2}-{x}^{2}{y}^{4}+2
\,\ln  \left( y \right) {x}^{2}{y}^{2}+2\,x{y}^{2}-1}}
\end{equation}

\begin{equation}
\label{harde4}
\frac{dy}{dx}=-{\frac {3\, \left( {{\rm e}^{y}} \right) ^{2}{x}^{4}{y}^{2}-7\,{
{\rm e}^{y}}{x}^{3}y+3\,{x}^{2}-y}{x \left(  \left( {{\rm e}^{y}}
 \right) ^{2}{x}^{4}{y}^{2}-3\,{{\rm e}^{y}}{x}^{3}y-{x}^{3}{{\rm e}^{
y}}+{x}^{2}-y-1 \right) }}
\end{equation}

\begin{equation}
\label{harde5}
\frac{dy}{dx}=-{\frac {4\,{x}^{3}{y}^{6}+8\,{x}^{4}{y}^{3}+4\,{x}^{5}-{x}^{4}+\cos
 \left( y \right) }{ \left( {y}^{6}+2\,x{y}^{3}+{x}^{2} \right) \sin
 \left( y \right) -3\,{x}^{4}{y}^{2}+3\,\cos \left( y \right) {y}^{2}}}
\end{equation}

\bigskip

\item  The following five 1ODEs have a non-elementary Liouvillian solution. We will refer to them as 1ODEs 6, 7, 8, 9, 10.
\begin{equation}
\label{harde6}
\frac{dy}{dx}={\frac { \left( 2\,{x}^{2}-y \right)  \left( \ln  \left( x \right) {y}
^{2}-{x}^{2}y-1 \right) y}{x \left(  \left( \ln  \left( x \right)
 \right) ^{2}{y}^{3}-\ln  \left( x \right) {x}^{2}{y}^{2}- \left( \ln
 \left( x \right)  \right) ^{2}{y}^{2}+2\,\ln  \left( x \right) {x}^{2
}y-{x}^{4}-\ln  \left( x \right) y \right) }}
\end{equation}

\begin{equation}
\label{harde7}
\frac{dy}{dx}=-{\frac { \left( \ln  \left( x \right)  \right) ^{2}x{y}^{2}-2\,\ln
 \left( x \right) {x}^{2}y-\ln  \left( x \right) xy+\ln  \left( x
 \right) {y}^{2}+{x}^{3}}{x \left( \ln  \left( x \right)  \right) ^{2}y}}
\end{equation}

\begin{equation}
\label{harde8}
\frac{dy}{dx}={\frac { \left( -{{\rm e}^{2\,x}}{x}^{6}{y}^{2}-2\,{{\rm e}^{2\,x}}{x}
^{5}{y}^{2}+{{\rm e}^{2\,x}}{x}^{4}{y}^{2}+{{\rm e}^{x}}{x}^{4}y+3\,{
{\rm e}^{x}}{x}^{3}y+1 \right) {{\rm e}^{-x}}}{{x}^{3} \left( {{\rm e}
^{x}}{x}^{3}y-x-1 \right) }}
\end{equation}

\begin{equation}
\label{harde9}
\frac{dy}{dx}=-{\frac { \left( \ln  \left( y \right)  \right) ^{2}x{y}^{4}- \left(
\ln  \left( y \right)  \right) ^{2}{y}^{4}-{x}^{2}\ln  \left( y
 \right) {y}^{2}+\ln  \left( y \right) x{y}^{2}+{x}^{3}}{xy \left( 2\,
\ln  \left( y \right) +1 \right)  \left( \ln  \left( y \right) {y}^{2}
-{x}^{2}-x \right) }}
\end{equation}

\begin{equation}
\label{harde10}
\frac{dy}{dx}={\frac {{y}^{3} \left( -x{y}^{3}+{{\rm e}^{y}}+1 \right) }{{{\rm e}^{y
}}{x}^{2}{y}^{6}+3\,{x}^{2}{y}^{5}-2\,{{\rm e}^{y}}x{y}^{3}-3\,{
{\rm e}^{y}}x{y}^{2}-3\,x{y}^{2}+{{\rm e}^{y}}}}
\end{equation}

\end{enumerate}

\bigskip

Applying our method (as implemented by us) each one of 1ODEs (ode[i]) will have the following solutions:

$$
sol\_ode[1] =-{\frac {\ln  \left( {x}^{2}-y \right) \ln  \left( x \right) {x}^{2}y+
\ln  \left( {x}^{2}-y \right) +1}{\ln  \left( x \right) {x}^{2}y+1}}={
\it \_C1}
$$
time consumed (in seconds)$$ 0.563 $$

\medskip

$$
sol\_ode[2] = -{\frac {x{y}^{2}{{\rm e}^{x}}+1}{{{\rm e}^{x}}\ln  \left( {x}^{3}-y
 \right) x{y}^{2}+\ln  \left( {x}^{3}-y \right) +y}}={\it \_C1}
$$
time consumed (in seconds)$$ 1.078 $$

\medskip

$$
sol\_ode[3] = \left( -{x}^{2}+\ln  \left( y \right)  \right) {{\rm e}^{{\frac {x}{x
{y}^{2}-1}}}}={\it \_C1}
$$
time consumed (in seconds)$$ 0.209 $$

\medskip

$$
sol\_ode[4] = -{\frac {{{\rm e}^{y}}xy\ln  \left( {x}^{3}{{\rm e}^{y}}+1 \right) -
\ln  \left( {x}^{3}{{\rm e}^{y}}+1 \right) +1}{xy{{\rm e}^{y}}-1}}={
\it \_C1}
$$
time consumed (in seconds)$$ 0.625 $$
\medskip

$$
sol\_ode[5] = \ln  \left( -2\,{{\rm e}^{iy}}{x}^{4}+{{\rm e}^{2\,iy}}+1 \right) +{
\frac {-i{y}^{4}-iyx+1}{{y}^{3}+x}}={\it \_C1}
$$
time consumed (in seconds)$$ 3.687 $$

\medskip

$$
sol\_ode[6] = - \frac{{\it Ei} \left( 1,- \left( \ln  \left( x \right) y-{x}^{2}
 \right) ^{-1} \right) y+{{\rm e}^{ \left( \ln  \left( x \right) y-{x}
^{2} \right) ^{-1}}}}{y}={\it \_C1}
$$
time consumed (in seconds)$$ 0.203 $$

\medskip

$$
sol\_ode[7] = \left( {\it Ei} \left( 1, \left( \ln  \left( x \right) y-x \right) ^{
-1} \right) {{\rm e}^{ \left( \ln  \left( x \right) y-x \right) ^{-1}}
}+x \right) {{\rm e}^{- \left( \ln  \left( x \right) y-x \right) ^{-1}
}}={\it \_C1}
$$
time consumed (in seconds)$$ 0.218 $$

\medskip

$$
sol\_ode[8] = \left( {\it Ei} \left( 1,- \left( {{\rm e}^{x}}{x}^{2}y-1 \right) ^{-
1} \right) x+{{\rm e}^{ \left( {{\rm e}^{x}}{x}^{2}y-1 \right) ^{-1}}}
 \right) {x}^{-1}={\it \_C1}
$$
time consumed (in seconds)$$ 0.687 $$

\medskip

$$
sol\_ode[9] = x\,{{\rm e}^{\displaystyle{\frac {x}{\ln  \left( y \right) {y}^{2}-x}}}}+{\it Ei}
 \left( 1,-{\frac {x}{\ln  \left( y \right) {y}^{2}-x}} \right) ={\it
\_C1}
$$
time consumed (in seconds)$$ 0.437 $$

\medskip

$$
sol\_ode[10] = \left( {{\rm e}^{- \left( x{y}^{3}-1 \right) ^{-1}}}{\it Ei} \left( 1
,- \left( x{y}^{3}-1 \right) ^{-1} \right) +{{\rm e}^{y}} \right) {
{\rm e}^{ \left( x{y}^{3}-1 \right) ^{-1}}}={\it \_C1}
$$
time consumed (in seconds)$$ 0.297 $$

\medskip
\medskip
\medskip

\begin{table}[h]
\begin{tabular}
{|c|c|c|}
\hline
 1ODE & Time (sec.) & Memory (MB) \\
\hline
 1 (\ref{harde1}) & 0,672 & 21,07\\
\hline
 2 (\ref{harde2}) & 1,078 & 35,60\\
\hline
 3 (\ref{harde3}) & 0,209 & 4,18\\
\hline
 4 (\ref{harde4}) & 0,625 & 33,18\\
\hline
 5 (\ref{harde5}) & 3,687 & 28,47\\
\hline
 6 (\ref{harde6}) & 0,329 & 17,41\\
\hline
 7 (\ref{harde7}) & 0,218 & 16,18\\
\hline
 8 (\ref{harde8}) & 0,687 & 21,26\\
\hline
 9 (\ref{harde9}) & 0,437 & 29,86\\
\hline
 10 (\ref{harde10}) & 0,297 & 18,42\\
\hline
\end{tabular}
\caption{Summary of time and memory consumption for the 10 1ODEs}
\end{table}
\medskip

\begin{obs}
\label{hardescomm}
Some computational comments:
\begin{itemize}
\label{computers}
    \item When we start a Maple session (the CAS we have used \footnote{One {\bf important} observation: for the purpose of this paper, it is nor relevant which CAS we have used to implement our method, could be any one of the reader's choice, it is not, strictly speaking, a computational paper. The theoretical results apply whichever Computer Algebra System one uses. For that matter, the qualitative results apply for any means of calculating the results}), the simple loading of the basic packages that are activated with the opening of the program's standard platform already uses 4.18 MB. So, when that number appears (in the table above), the meaning is basically `very low memory expenditure'.

    \item For the measurement of consumed CPU time, a similar problem occurs: the time measurement provided is `variable from session to session'. However, the number representing the time spent is always reasonably close to the average of the values obtained in different rounds. In the data placed in the table above, we avoid averaging over many sessions (which means that the last two digits are not significant) because we were focused on a more qualitative result.



    \item Finally, looking at the table below, we can see why these 1ODEs can present difficulties in order to be solved. We only need to pay attention to the Lie symmetries or to the integrating factors of the 1ODEs' set. We see that only 1ODEs 3, 5 and 10 have integrating factors in which the $\theta$ function is absent (and, even so, they are not simple).
\end{itemize}
\end{obs}

\medskip

\begin{tabular}
{|c|c|c|}
\hline
 1ODE & Symmetry ($\nu$) & Integrating Factor \\
\hline
 1 (\ref{harde1}) & $-{\frac { ( {x}^{2}y\ln  \left( x \right) +1 ) ^{2}}{{
{\rm e}^{ \left( {x}^{2}y\ln  \left( x \right) +1 \right) ^{-1}}}
 \left(  \left( \ln  \left( x \right)  \right) ^{2}{x}^{4}{y}^{2}+\ln
 \left( x \right) {x}^{4}+{x}^{2}y\ln  \left( x \right) +1 \right) }}
$ & ${\frac {{{\rm e}^{ ( {x}^{2}y\ln  \left( x \right) +1 ) ^{-1}}}}{ \left( {x}^{2}y\ln  \left( x \right) +1 \right) ^{2}}}$\\
\hline
 2 (\ref{harde2}) & $- \frac{ ( x{y}^{2}{{\rm e}^{x}}+1 ) ^{2} \, {{\rm e}^{{\frac
{-y}{x{y}^{2}{{\rm e}^{x}}+1}}}}  }{  \left( {{\rm e}^
{x}} \right) ^{2}{x}^{2}{y}^{4}+{{\rm e}^{x}}{x}^{4}{y}^{2}-{{\rm e}^{
x}}x{y}^{3}+2\,x{y}^{2}{{\rm e}^{x}}-{x}^{3}+y+1 }
$ & $\frac{{\rm e}^{{\frac {y}{x{y}^{2}{{\rm e}^{x}}+1}}}}{ \left( x{y}^{2}{{\rm e}^{x}}+1 \right) ^{2}}$ \\
\hline
 3 (\ref{harde3}) & $ \frac{ ( x{y}^{2}-1 ) ^{2}y \, {{\rm e}^{{\frac {-x}{x{y}^{2}-
1}}}} }{ -2\,{x}^{4}{y}^{2}-{x}^{2}{y}^{4}+2\,\ln
 \left( y \right) {x}^{2}{y}^{2}+2\,x{y}^{2}-1 }
$ & $- \frac{{\rm e}^{{\frac {x}{x{y}^{2}-1}}}}{y\, \left( x{y}^{2}-1 \right) ^{2}}$ \\
\hline
 4 (\ref{harde4}) & ${\frac { \left( xy{{\rm e}^{y}}-1 \right) ^{2}\,{{\rm e}^{ \left(1- xy{
{\rm e}^{y}} \right) ^{-1}}}}{{{\rm e}^{y}}x \left(  \left( {{\rm e}^{
y}} \right) ^{2}{x}^{4}{y}^{2}-3\,{{\rm e}^{y}}{x}^{3}y-{x}^{3}{
{\rm e}^{y}}+{x}^{2}-y-1 \right) }}
$ & ${\frac {-{{\rm e}^{ \left( xy{{\rm e}^{y}}-1 \right) ^{-1}}}{{\rm e}^{y}}}{ \left( xy{{\rm e}^{y}}-1 \right) ^{2}}}$ \\
\hline
 5 (\ref{harde5}) & ${\frac { ( {y}^{3}+x )^{2}\,{{\rm e}^{ ( -{y}^{3}-x
) ^{-1}}}}{ \left( \sin \left( y \right) {y}^{6}-3\,{x}^{4}{y}^{2
}+2\,\sin \left( y \right) x{y}^{3}+3\,\cos \left( y \right) {y}^{2}+
\sin \left( y \right) {x}^{2} \right) }}
$ & ${\frac {-{{\rm e}^{ ( {y}^{3}+x ) ^{-1}}}}{ \left( {y}^{3}+x \right) ^{2}}}$ \\
\hline
 6 (\ref{harde6}) & ${\frac { ( \ln  \left( x \right) y-{x}^{2} ) ^{2}{y}^{2}\,{{\rm e}^{({x}^{2} -  \ln  \left( x \right) y) ^{-1}}}}{
  \left( \ln  \left( x \right)  \right) ^{2}{y}^{3}-\ln  \left( x \right) {x}^{2}{y}^{2}- \left( \ln  \left( x \right)
 \right) ^{2}{y}^{2}+2\,\ln  \left( x \right) {x}^{2}y-{x}^{4}-\ln
 \left( x \right) y  }}
$ & ${\frac {-{{\rm e}^{ ( \ln  \left( x \right) y-{x}^{2}) ^{-1}}}}{ \left( \ln  \left( x \right) y-{x}^{2} \right) ^{2}x{y}^{2}}}$ \\
\hline
 7 (\ref{harde7}) & $\frac{ ( \ln  \left( x \right) y-x ) ^{2} }{ {{\rm e}^{-
 \left( \ln  \left( x \right) y-x \right) ^{-1}}} \, ( \ln  \left( x \right)  ) ^{2}\,y}
$ & $\frac{-{\rm e}^{- \left( \ln  \left( x \right) y-x \right) ^{-1}}}{(
\ln  \left( x \right) y-x ) ^{2}\,x}$ \\
\hline
 8 (\ref{harde8}) & $\frac{ {{\rm e}^{-{\frac {{{\rm e}^{x}}{x}^{3}y-x+1}{{{\rm e}^{x}}{x}^{2}y-1}
}}} ( {{\rm e}^{x}}{x}^{2}y-1 ) ^{2} }{ {x}\, \left( {
{\rm e}^{x}}{x}^{3}y-x-1 \right) }
$ & $\frac{-{{\rm e}^{{\frac {{{\rm e}^{x}}{x}^{3}y-x+1}{{{\rm e}^{x}}{x}^{2}y-1}
}}} }{ {x}^{2} \left( {{\rm e}^{x}}{x}^{2}y-1 \right) ^{2}}$ \\
\hline
 9 (\ref{harde9}) & $ \frac{ {{\rm e}^{-{\frac {x}{\ln  \left( y \right) {y}^{2}-x}}}} ( \ln
 \left( y \right) {y}^{2}-x ) ^{2} }{ {y}\, \left( 2\,\ln
 \left( y \right) +1 \right)\,\left( \ln  \left( y \right) {y}^{2
}-{x}^{2}-x \right) }
$ & $ \frac{ -{{\rm e}^{{\frac {x}{\ln  \left( y \right) {y}^{2}-x}}}} }{ \left( \ln  \left( y \right) {y}^{2}-x \right) ^{2}\,{x} }$ \\
\hline
 10 (\ref{harde10}) & $\frac{ {{\rm e}^{- ( x{y}^{3}-1 ) ^{-1}}} ( x{y}^{3}-1 ) ^{2} }{ {{\rm e}^{y}}{x}^{2}{y}^{6}+3\,{x}^{2}{y}^{5}-2\,
{{\rm e}^{y}}x{y}^{3}-3\,{{\rm e}^{y}}x{y}^{2}-3\,x{y}^{2}+{{\rm e}^{y}}  }
$ & ${\frac {-{{\rm e}^{ ( x{y}^{3}-1 ) ^{-1}}}}{ \left( x{y}^{3}-1 \right) ^{2}}}$ \\
\hline
\end{tabular}

\begin{obs}
For all 1ODEs in Kamke's book \cite{Kam} that obey the conditions of application of the algorithms presented above our method does very well in all cases. As they are much simpler cases than those shown in the table, we decided not to include them in order not to increase the presentation unnecessarily.
\end{obs}


\medskip

\subsection{Integrability analisys}
\label{ub}
\hspace\parindent
In this section, we will present a very useful way to apply the method: if the 1ODE presents parameters the algebraic procedure `behind' the method brings us the possibility of an analysis of the integrability region of the 1ODE's parameters. To see a real case of using this to search for integrability regions in terms of Liouvillian functions (closed solutions), let's take as an example an 1ODE obtained from the study of 2ODEs that model nonlinear oscillation phenomena. In our case, the 1ODE we are going to present appears in the reduction of a Levinson-Smith equation\footnote{The Levinson-Smith equation has applications in astrophysics \cite{Ran}.} \cite{LeSm}
\begin{equation}
\label{levsmi}
\ddot{x} + F(x,\dot{x})\,\dot{x} + G(x) = 0,
\end{equation}
a 2ODE that is a generalization of the Liénard equation\footnote{There is a vast literature on the use of this equation to model various phenomena: ranging from electrical circuits, study of heartbeat, neuron activity, chemical kinetics to turbulence phenomena \cite{ Pol, PoMa, Fit, Str}.} \cite {Liena}
\begin{equation}
\label{lien}
\ddot{x} + F(x)\,\dot{x} + G(x) = 0.
\end{equation}
As 2ODE (\ref{levsmi}) does not explicitly depend on time, we can do the following coordinate transformation to reduce its order:
\begin{equation}
\label{tlevsmi}
\left\{ \begin{array}{l}
x = x, \\
\dot{x} = y, \\
\ddot{x}= \frac{dy}{dx} \, y.
\end{array} \right.
\end{equation}
The application of this transformation to the 2ODE (\ref{levsmi}) leads to the 1ODE given by
\begin{equation}
\label{1levsmi}
\frac{dy}{dx}= - \frac{F(x,y)\,y + G(x)}{y}.
\end{equation}  
We can now, from very generic $ F $ and $ G $ functions, try to obtain integrable cases in terms of Liouvilian functions (i.e., closed form solutions). Let's analyze the following case:
\begin{equation}
\label{levsminos}
\frac{dy}{dx}= {\frac {(-f{x}^{2}{{\rm e}^{x}}-b\,x{{\rm e}^{x}}-c\,{y}-a)\,y-e\,{x}^{2}{{\rm e}^{x}}-d\,x}{y}}
\end{equation}  
Adding the parameters $\{a,b,c,d,e,f\}$ to the coefficients to be determined we obtain that, if the parameters $d$ and $e$ are zero, we have a S-function  given by $ S= -{\frac {cz}{cy+a}}$. The 1ODE
\begin{equation}
\label{newde}
{\frac {d}{dx}}y \left( x \right) ={\frac {-f{x}^{2}y \left( x
 \right) {{\rm e}^{x}}-x{{\rm e}^{x}}y \left( x \right) b- \left( y
 \left( x \right)  \right) ^{2}c-y \left( x \right) a}{y \left( x
 \right) }}
\end{equation}
can be solved with the method:
$$
y  = \left( -\frac {f\,{\rm e}^x \left( \left( c+1 \right) ^{2}{x}^{2}-2\, \left( c+1 \right) x+2 \right) }{ \left( c+1 \right) ^3}-
{\frac {b\,{\rm e}^x \left(  \left( c+1 \right) x-1 \right) }{ \left( c+1 \right) ^{2}}}-{\frac {a}{c}}+{\it \_C1} \right) 
$$

\noindent
Replacing $ y $ with $ \dot{x} $, we obtain the first integral of the Levinson-Smith equation (for the parameter regions we find):
$$ I = \dot{x} +\left( \frac {f\,{\rm e}^x \left( \left( c+1 \right) ^{2}{x}^{2}-2\, \left( c+1 \right) x+2 \right) }{ \left( c+1 \right) ^3}+
{\frac {b\,{\rm e}^x \left(  \left( c+1 \right) x-1 \right) }{ \left( c+1 \right) ^{2}}}+{\frac {a}{c}}\right) 
$$

\section{Conclusion}
\label{conclu}
\hspace\parindent

Here, we have developed a new method to solve 1ODEs presenting an elementary function in their definition. Our procedure is applicable to 1ODEs presenting a Liouvillian general solution. Our approach deals with 1ODEs that can be expressed in the following form:
\begin{equation}
\label{1edoconclu}
y' = \frac{dy}{dx} = \phi(x,y,\theta) = \frac{M(x,y,\theta)}{N(x,y,\theta)}
\end{equation}
where $\phi $ is a rational function of $ (x, y, \theta) $ and $ \theta $ is an exponential or logarithmic function of $(x,y)$ ($ M $ and $ N $ are coprime polynomials in $ (x, y, \theta) $).

It is worth mentioning that every trigonometric function can be expressed as a combination of exponential functions. Actually, one such function, since, for example, one can always write $cos(f(x,y))$ as:

$$cos(f(x,y)) = (e^{if(x,y)} + e^{-if(x,y)}) / 2$$
$$cos(x) = ((e^{if(x,y)})^2 + 1) / e^{if(x,y)}$$
We can also do something analogous for each and every trigonometric functions. The same scenario applies to hyperbolic trigonometric functions, this time with exponential functions with non-imaginary arguments.

For the case of inverse trigonometric functions (also for the inverse hyperbolic ones) they can be expressed as a function of logarithms, So, all these functions are also covered by our approach as one can see an example (example 5) on section \ref{performance}, and are included when we say that {\bf $\theta$ is an exponential function} in the general expression for our 1ODE under scrutiny (eq. \ref{1edoconclu} above).

The method is based on a connection between an 1ODE with an elementary function $\theta$ and a polynomial vector field in three variables. The main point is that, if this 1ODE has a Liouvillian general solution, we can associate a rational 2ODE with the vector field and, therefore, use the $S$-function method (see \cite{Noscpc2019}) to find a first integral for the vector field and (using the simple substitution $z \rightarrow \theta$) find the general solution of the 1ODE.

One might consider that exchanging the problem of solving a 1ODE for the problem of finding a first integral of a 2ODE may not seem like a good idea at first. However, there are two facts that make this exchange result in a very efficient procedure: first, the 1ODE has an elementary function whereas the 2ODE is rational; second, the $S$-function method is effective in cases where the search for first integrals of 2ODEs by procedures such as the Lie symmetries method and Darboux approaches are particularly problematic.

The method we developed first ($ Lsolv $ algorithm) is very efficient compared to the more canonical alternative that would be to try a Darboux-Prelle-Singer (DPS) approach or the Lie symmetry method. However, the appearance of an unexpected result, improved the efficiency of our algorithm far surpassing our expectations. These result (expressed in the theorems \ref{theoAimp} and \ref{theoSsis}) allowed the creation of the $ FastL_S $ algorithm that, for a very comprehensive range of 1EDOs (the class $ L_S $ - see the definition \ref{lsset}), performed very well.


Although this paper is not, strictly speaking, a computational paper, the algorithms $ Lsolv $ and $ FastL_S $ can be implemented in order to make it more broadly available. We have actually implemented them in a computational package ($LeapS1ode$) on the Maple symbolic computing platform in order to produce the``computational" results, such as times of processing and memory used, testing the efficiency of the algorithms, etc. 

Finally, due to the nature of the algorithm search process, we can very straightforwardly perform  analysis and determination of integrability regions for any 1ODEs that present parameters (see the example in section \ref{ub}). So our whole approach is very suitable for dealing with chaotic systems and solving them.



\begin{thebibliography}{25}

\bibitem{Noscpc2002}  L.G.S. Duarte, S.E.S.Duarte, L.A.C.P. da Mota and J.F.E.
Skea, {\it Extension of the Prelle-Singer Method and a MAPLE
implementation}, Computer Physics Communications, Holanda, v. 144, n. 1, p. 46-62 (2002).

\bibitem{Noscpc2012}
J.Avellar, L.G.S.DuarteL.A.C.P.da Mota, PSsolver: A Maple implementation to solve first order ordinary differential equations with Liouvillian solutions,
Computer Physics Communications
Volume 183, Issue 10, October 2012, Page 2313

\bibitem{Noscpc2019}
J. Avellar, M.S. Cardoso, L.G.S. Duarte, L.A.C.P. da Mota
{\it Dealing with Rational Second Order Ordinary Differential Equations where both Darboux and Lie Find It Difficult: The S-function Method},
Computer Physics Communications, {\bf 234}, (2019) 302-314.

\bibitem{astro}
J.-M. Huré, F. Hersant
{\it A new equation for the mid-plane potential of power law disks}
 Volume 467 / No 3 (June I 2007)  Astronomy \& Astrophysica (A\&A), 467 3 (2007) 907-910

\bibitem{Dar}
G. Darboux,
{\it M\'emoire sur les \'equations diff\'erentielles alg\'ebriques du premier ordre et du premier
degr\'e (M\'elanges)},
Bull. Sci. Math. 2\`eme s\'erie 2, 60-96, 2, 123-144, 2, 151-200 (1878).

\bibitem{PrSi}
M. Prelle and M. Singer,
Elementary first integral of differential equations.
{\it Trans. Amer. Math. Soc.}, {\bf  279} 215 (1983).

\bibitem{Sin}
M. Singer,
{\it Liouvillian First Integrals},
Trans. Amer. Math. Soc., {\bf  333} 673-688 (1992).

\bibitem{Chr} C. Christopher,
{\it Liouvillian first integrals of second order polynomial differential equations},
Electron. J. Differential Equations, {\bf 49}, (1999), 7 pp. (electronic).

\bibitem{CaLl}
L. Cairó and J. Llibre,
{\it Darboux Integrability for 3D Lotka-Volterra systems},
J. Phys. A: Math. Gen., {\bf 33} 2395-2406 (2000).

\bibitem{ChLlPaZh} C. Christopher, J. Llibre, C. Pantazi and X. Zhang,
{\it Darboux integrability and invariant algebraic curves for planar polynomial systems},
J. Phys. A, {\bf 35}, (2002) 2457–2476. https://doi.org/10.1088/0305-4470/35/10/310

\bibitem{ChGiGiLl} J. Chavarriga, H. Giacomini, J. Giné and J. Llibre,
{\it Darboux integrability and the inverse integrating factor},
J. Differential Eqs., {\bf 194}, (2003) 116–139. https://doi.org/10.1016/S0022-0396(03)00190-6

\bibitem{Nosjcam2005} J. Avellar, L.G.S. Duarte, S.E.S. Duarte, L.A.C.P. da
Mota, {\it Integrating First-Order Differential Equations with
Liouvillian Solutions via Quadratures: a Semi-Algorithmic Method,}
Journal of Computational and Applied Mathematics {\bf 182}, 327-332, (2005).

\bibitem{ChLlPe} C. Christopher, J. Llibre and J.V. Pereira,
{\it Multiplicity of Invariant Algebraic Curves in Polynomial Vector Fields},
Pacific Journal of Mathematics, {\bf 229}, (2007) 63-117. https://doi.org/10.2140/pjm.2007.229.63

\bibitem{ChLlPaWa2} C. Christopher, J. Llibre, C. Pantazi and S. Walcher,
{\it Inverse problems for invariant algebraic curves: Explicit computations},
Proc. Roy. Soc. Edinburgh, {\bf 139}, (2009) 287-302.  https://doi.org/10.1017/S0308210507001175

\bibitem{ChLlPaWa3} C. Christopher, J. Llibre, C. Pantazi and S. Walcher, {\it Darboux integrating factors: inverse problems},
J. Differential Equations, {\bf 250}, (2011) 1-25. https://doi.org/10.1016/j.jde.2010.10.013

\bibitem{Zha} X. Zhang,
{\it Liouvillian integrability of polynomial differential systems},
Trans. Amer. Math. Soc., {\bf 368}, (2016) 607-620. https://doi.org/10.1090/S0002-9947-2014-06387-3

\bibitem{FeGi} A. Ferragut and H. Giacomini,
{\it A New Algorithm for Finding Rational First Integrals of Polynomial Vector Fields},
Qual. Theory Dyn. Syst., {\bf 9}, (2010) 89–99. https://doi.org/10.1007/s12346-010-0021-x

\bibitem{BoChClWe}  A. Bostan, G. Chèze, T. Cluzeau and J.-A. Weil,
{\it Efficient algorithms for computing rational first integrals and Darboux polynomials of planar polynomial vector fields},
Mathematics of Computation, {\bf 85}, (2016) 1393-1425. https://doi.org/10.1090/mcom/3007

\bibitem{ChCo}  G. Chèze and T. Combot,
{\it Symbolic Computations of First Integrals for Polynomial Vector Fields},
Foundations of Computational Mathematics, (2019). https://doi.org/10.1007/s10208-019-09437-9

\bibitem{Olv}
P.J. Olver,
{\it Applications of Lie Groups to Differential Equations},
Springer-Verlag, (1986).

\bibitem{Ibr} N.H. Ibragimov,
{\it Elementary Lie Group Analysis and Ordinary Differential Equations},
Wiley: Chichester, (1999).

\bibitem{BlAn}
G.W. Bluman and S.C. Anco,
{\it Symmetries and Integration Methods for Differential Equations},
Applied Mathematical Series {\bf Vol. 154}, Springer-Verlag, New York, (2002).

\bibitem{Sch}
F. Schwarz,
{\it Algorithmic Lie Theory for Solving Ordinary Differential Equations},
Chapman  Hall / CRC -- Taylor and Francis Group, (2008).

\bibitem{Noscpc1997} E.S. Cheb-Terrab, L.G.S. Duarte and L.A.C.P. da Mota,
{\it Computer Algebra Solving of First Order ODEs Using Symmetry Methods}.
Comput.Phys.Commun., {\bf 101}, 254, (1997).

\bibitem{Noscpc1998} E.S. Cheb-Terrab, L.G.S. Duarte and L.A.C.P. da Mota,
{\it Computer Algebra Solving of Second Order ODEs Using Symmetry Methods}.
Comput.Phys.Commun., {\bf 108}, 90, (1998).

\bibitem{AbGu}
B. Abraham-Shrauner and A. Guo,
{\it Hidden Symmetries Associated with the Projective
Group of Nonlinear First-Order Ordinary Differential Equations}.
J. Phys. A: Math.Gen., {\bf 25}, 5597-5608, (1992).

\bibitem{GoLe} K.S. Govinder and P.G.L Leach,
{\it A group theoretic approach to a class of second-order ordinary differential
equations not possesing Lie point symmetries}.
J. Phys. A: Math. Gen., {\bf 30}, 2055-68, (1997).

\bibitem{AdMa} A.A. Adam and F.M. Mahomed,
{\it Non-local symmetries of first-order equations}.
IMA J. Appl. Math., {\bf 60}, 187-98, (1998).

\bibitem{GaBrSe} M.S. Bruzón, M.L. Gandarias and M. Senthilvelan,
{\it Nonlocal symmetries of Riccati and Abel chains and their similarity reductions}.
 Journal of Mathematical Physics, {\bf 53}, 023512 (2012).

\bibitem{MuRo}
C. Muriel and J.L. Romero,
{\it New methods of reduction for ordinary differential equations},
IMA J. Appl. Math., {\bf 66}(2), 111-125, (2001).

\bibitem{MuRo2}
C. Muriel and J.L. Romero,
{\it $\,C^{\,\infty}$-Symmetries and reduction of equations without Lie
point symmetries},
J. Lie Theory, {\bf 13}(1), 167-188, (2003).

\bibitem{PuSa}
E. Pucci and G. Saccomandi,
{\it On the reduction methods for ordinary differential equations}.
J. Phys. A: Math. Gen., {\bf 35}, 6145-6155, (2002).

\bibitem{Nuc}
M.C. Nucci,
{\it Jacobi Last Multiplier and Lie Symmetries:
A Novel Application of an Old Relationship},
Journal of Nonlinear Mathematical Physics, {\bf 12}(2), 284-304, (2005).

\bibitem{Nosjpa2002-1} L.G.S. Duarte, S.E.S.Duarte and L.A.C.P. da Mota,
{\it A method to tackle first order ordinary differential equations with
Liouvillian functions in the solution}, in J. Phys. A: Math. Gen.,
{\bf 35} 3899-3910 (2002).

\bibitem{Nosjpa2002-2} L.G.S. Duarte, S.E.S.Duarte and L.A.C.P. da Mota,
{\it Analyzing the Structure of the Integrating Factors for First Order
Ordinary Differential Equations with Liouvillian Functions in the
Solution}, J. Phys. A: Math. Gen., {\bf 35} 1001-1006 (2002).

\bibitem{Nosamc2007}
J. AvellarL.G.S. Duarte, S.E.S.Duarte and L.A.C.P. da Mota,
{\it A semi-algorithm to find elementary first order invariants
of rational second order ordinary differential equations}, Appl. Math. Comp., {\bf
184} 2-11 (2007).

\bibitem{Nosjpa2001}
L.G.S. Duarte, S.E.S.Duarte, L.A.C.P. da Mota and J.F.E. Skea,
{\it Solving second order ordinary differential equations by
extending the Prelle-Singer method}, J. Phys. A: Math.Gen., {\bf
34} 3015-3024 (2001).

\bibitem{LaRa}
M. Lakshmanan and S. Rajasekar, {\it Nonlinear Dynamics:
Integrability, Chaos and Patterns}. New York: Springer-Verlag (2003).

\bibitem{ChSeLa}
V.K. Chandrasekar, M. Senthilvelan and M. Lakshmanan, {\it On the
complete integrability and linearization of certain second order
nonlinear ordinary differential equations}. Proceedings of the Royal
Society London Series A, {\bf 461}, Number 2060, 2005.

\bibitem{Nosjmp2009}
L.G.S.Duarte and L.A.C.P.da Mota,
{\it Finding Elementary First Integrals for Rational Second Order Ordinary Differential Equations},
J. Math. Phys., {\bf 50}, (2009).

\bibitem{Nosjpa2010}
L.G.S.Duarte and L.A.C.P.da Mota,
{\it 3D polynomial dynamical systems with elementary first integrals},
J. Phys. A: Math. Theor. {\bf 43}, n.6, (2010). doi:10.1088/1751-8113/43/6/065204

\bibitem{GoPiSe}
P. R. Gordoa, A. Pickering and M. Senthilvelan,
{\it The Prelle-Singer method and Painleve hierarchies},
J. Math. Phys., {\bf 55}, 053510 (2014)

\bibitem{TPCSL}
A. K. Tiwari, S. N. Pandey, V. K. Chandrasekar, M. Senthilvelan and M. Lakshmanan,
{\it The inverse problem of a mixed Liénard-type nonlinear oscillator equation from symmetry perspective},
Acta Mechanica, {\bf 227}, Issue 7, (2016) 2039–2051.

\bibitem{MCSL}
R. Mohanasubha, V. K. Chandrasekar, M. Senthilvelan, M. Lakshmanan
{\it Interplay of symmetries and other integrability quantifiers in finite-dimensional integrable nonlinear dynamical systems}
Proc. R. Soc. A., {\bf 472}: 220150847 (2016). http://doi.org/10.1098/rspa.2015.0847

\bibitem{tito1} C Muriel and J L Romero
First integrals, integrating factors and $\lambda$-symmetries of second-order differential equations
Journal of Physics A: Mathematical and Theoretical, {\bf 42}, Number 36

\bibitem{tito2}
C Muriel and J L Romero
Contribution to the Special Issue “Symmetries of Differential Equations: Frames, Invariants and Applications”
Nonlocal Symmetries, Telescopic Vector Fields and $\lambda$-Symmetries of Ordinary Differential Equations
SIGMA 8 (2012), 106

\bibitem{Noscpc2007} J. Avellar, L.G.S. Duarte, S.E.S.Duarte and L.A.C.P. da Mota,
{\it Determining Liouvillian first integrals for dynamical systems in the plane},
Computer Physics Communications, {\bf 177}, (2007) 584-596. https://doi.org/10.1016/j.cpc.2007.05.014

\bibitem{Nosjde2021} L.G.S. Duarte and L.A.C.P. da Mota,
{\it An efficient method for computing Liouvillian first integrals of planar polynomial vector fields},
Journal of Differential Equations, {\bf 300}, (2021) 356-385. https://doi.org/10.1016/j.jde.2021.07.045

\bibitem{Dav}
J.H. Davenport, Y. Siret and E. Tournier,
{\it Computer Algebra: Systems and Algorithms for Algebraic Computation}.
Academic Press, Great Britain (1993).

\bibitem{Kam}
E. Kamke,
{\it Differentialgleichungen: L{\"o}sungsmethoden und L{\"o}sungen},
Chelsea Publishing Co, New York (1959).

\bibitem{LeSm}
N. Levinson and O. Smith,
{\it A general equation for relaxation oscillations},
Duke Mathematical Journal, {\bf 9}, (1942), 382-403.

\bibitem{Ran}
Z. Ran,
{\it One exactly soluble model in isotropic turbulence},
Advances and Applications in Fluid Mechanics, {\bf 5}, (2009), 41-47.

\bibitem{Liena}
A. Liénard,
{ Revue générale de l'électricité},
{\bf 23}, (1928), 901- 912, 946-954.

\bibitem{Pol}
B. van der Pol,
{\it On relaxtion-oscillations},
The London, Edinburgh and Dublin Philosophical Magazine and Journal of Science, {\bf 2}, (1927) 978-992.

\bibitem{PoMa}
B. van der Pol and J. van der Mark,
{\it The heart beat considered as a relaxation oscillations and an electrical model of the heart},
The London, Edinburgh and Dublin Philosophical Magazine and Journal of Science, {\bf 6}, (1928), 763-775.

\bibitem{Fit}
F. Fitzhugh,
{\it Impulses and physiological states in theoretical models of nerve membranes},
Biophysics Journal, {\bf 1}, (1961), 445-466.

\bibitem{Str}
S. H. Strogatz,
{\it Nonlinear Dynamics and Chaos},
Addison-Wesley, Reading, Massachussets, (1994).




\end{thebibliography}
\end{document}